\documentclass{pasj00}
\draft

\usepackage{graphicx,times,psfig}

\begin{document}

\title{{\it Suzaku} Observations of the cluster of galaxies Abell~2052 (accepted PASJ, 2008)}

\author{
Takayuki \textsc{Tamura}\altaffilmark{1},
Kazuhisa \textsc{Mitsuda}\altaffilmark{1}, 
Yoh \textsc{Takei}\altaffilmark{1,2}, 
Noriko Y. \textsc{Yamasaki}\altaffilmark{1}, \\
Akiharu \textsc{ITOH}\altaffilmark{1}, 
Kiyoshi \textsc{Hayashida}\altaffilmark{3},
J. Patrick \textsc{Henry}\altaffilmark{4},
Hideyo \textsc{Kunieda}\altaffilmark{5}, \\
Kyoko \textsc{Matsushita}\altaffilmark{6},
Takaya \textsc{Ohashi}\altaffilmark{7},
}
\altaffiltext{1}{
Institute of Space and Astronautical Science,
Japan Aerospace Exploration Agency,\\
3-1-1 Yoshinodai, Sagamihara, Kanagawa 229-8510
}
\email{ttamura@isas.jaxa.jp}
\altaffiltext{2}{SRON Netherlands Organization for Space Research,
Sorbonnelaan 2, 3584CA, Utrecht, The Netherlands}

\altaffiltext{3}{Department of Astrophysics, Faculty of Science, Osaka University, Toyonaka 560-0043}
\altaffiltext{4}{Institute for Astronomy, University of Hawai'i, 2680 Woodlawn Drive, Honolulu, HI 96822, USA}
\altaffiltext{5}{Department of Particle and Astrophysics, Nagoya University, Furo-cho, Chikusa-ku, Nagoya 464-8602}
\altaffiltext{6}{Department of Physics, Tokyo University of Science, 1-3 Kagurazaka, Shinjuku, Tokyo 162-8601}
\altaffiltext{7}{Department of Physics, Tokyo Metropolitan University,
1-1 Minami-Osawa, Hachioji, Tokyo 192-0397}
\KeyWords{
cosmology: large-scale structure ---
galaxies: clusters: individual (A2052) ---
intergalactic medium ---
X-rays: diffuse background
}

\maketitle

\begin{abstract}
The results from {\it Suzaku} XIS observations of the relaxed cluster of galaxies Abell~2052 are presented.
Offset pointing data are used to estimate the Galactic foreground emission in the direction to the cluster.
Significant soft X-ray excess emission above this foreground, the intra-cluster medium emission, and other background components
is confirmed and resolved spectroscopically and radially.
This excess can be described either by (a) local variations of known Galactic emission components 
or (b) an additional thermal component with temperature of about 0.2~keV, possibly associated with the cluster.
The radial temperature and metal abundance profiles of the intra-cluster medium are measured within $\sim 20'$ in radius (about 60\% of the virial radius) from the cluster center .
The temperature drops radially to $0.5-0.6$ of the peak value at a radius of $\sim 15'$.
The gas-mass-weighted metal abundance averaged over the observed region is found to be $0.21\pm0.05$ times solar.
\end{abstract}

\section{Introduction}
In a limited number of clusters, soft X-ray excess emission above that predicted from the intra-cluster medium (ICM) contribution has been discovered (e.g. Kaastra et al. 2003a).
This phenomenon possibly originates from a 
warm-hot intergalactic medium (WHIM) associated with the cluster.
This not-yet identified WHIM at a temperature of $10^5-10^7$~K 
is predicted to reside and constitute 
about half of the total baryon mass at low redshift 
in cosmic simulations in a $\Lambda$CDM universe
(e.g. Cen \& Ostriker 1999).

In spite of a number of investigations,
no soft excess feature has been robustly identified as the 
WHIM {\it associated with the cluster}.
These identifications have been challenging 
largely because of bright and diffuse foreground contamination.
This foreground emission includes separate origins such as the local hot bubble and the Galactic halo
toward which different absorptions should be applied to.
Therefore its brightness varies from sky to sky in a complex manner.
Furthermore, the spectral nature of the WHIM may be similar to that of the Galactic foreground.
To separate the emission associated with the cluster from the Galactic one and other possibilities,
sensitivities of previous measurements were limited. 
Bregman (2007) reviewed the current observational situation in detail.

In a search for the nature of soft excess emission,
we have observed the Abell 2052 cluster of galaxies (A~2052; redshift $z=0.0356$) with 
the {\it Suzaku} X-ray Imaging Spectrometer (XIS).
This is one of the X-ray brightest and nearest objects among clusters showing a significant soft excess (Kaastra et al. 2003a).
The XIS has a good sensitivity in the soft X-ray band for extended sources.
We could resolve the soft excess emission in the outer part of the cluster
both spatially and spectroscopically.
This analysis in turn provided accurate measurements of temperature and metallicity radial profiles within a radius of $\sim 20'$ or 800~kpc.

Throughout this paper,
we assume cosmological parameters as follows; $H_0 = 70$ km s$^{-1}$Mpc$^{-1}$, 
$\Omega_\mathrm{m} = 0.3$, and $\Omega_\mathrm{\Lambda} = 0.7$.
One arc-minute corresponds to 42~kpc at the cluster distance.
We use the 90\% confidence level unless stated otherwise.

\section{Observations}\label{sect:obs}
\begin{figure}[hp]
\begin{center}
\leavevmode\psfig{figure=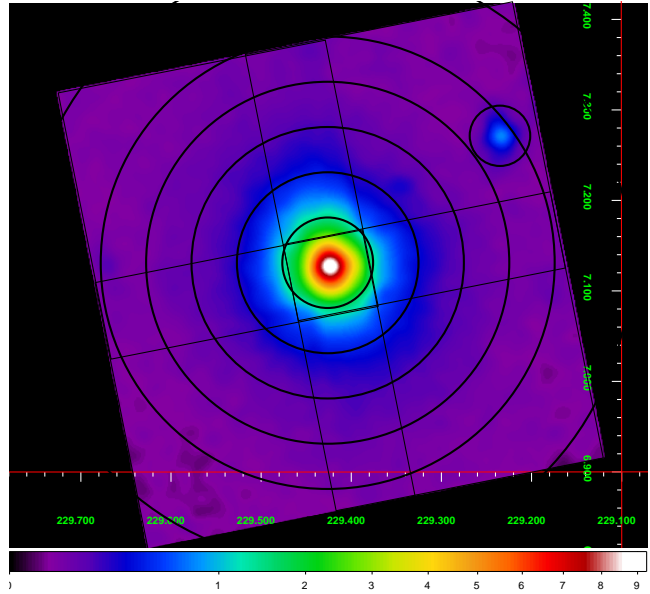,width=85mm}
  \end{center}
  \caption{Mosaic {\it Suzaku} image of A~2052 in the 1--3 keV band. 
The vignetting was corrected, but no background was subtracted. 
The spectra extracted regions (radii of $3', 6', 9', 12', 15', 20'$) are shown in circles. 
A contamination source is marked by a circle at upper right.
The XIS field of view ($17' \times 17'$ square) of the four pointings are also shown 
in lines.
}
\label{fig:xisimage}
\end{figure}

{\it Suzaku} observations of A~2052 were performed as a part of performance verifications.
To cover a large area around the cluster,
we performed four pointings as shown in Table.~\ref{tbl:obs}.
In each pointing, the cluster center was on the CCD corner.
Hereafter we refer these pointings as P1, P2, P3 and P4.
This results in a $30' \times 30'$ square field with the XIS.
Figure ~\ref{fig:xisimage} shows a mosaic X-ray image of the cluster.
To estimate the Galactic foreground emission we also performed a offset pointing, 
with an angle \timeform{3D.74} from the cluster center.
This offset position includes no strong X-ray source
and is not only close enough to have a similar level of the Galactic foreground
but also far away from the cluster and other associated clusters
to avoid the clusters emission.

\begin{figure}[hp]
\begin{center}
\leavevmode\psfig{figure=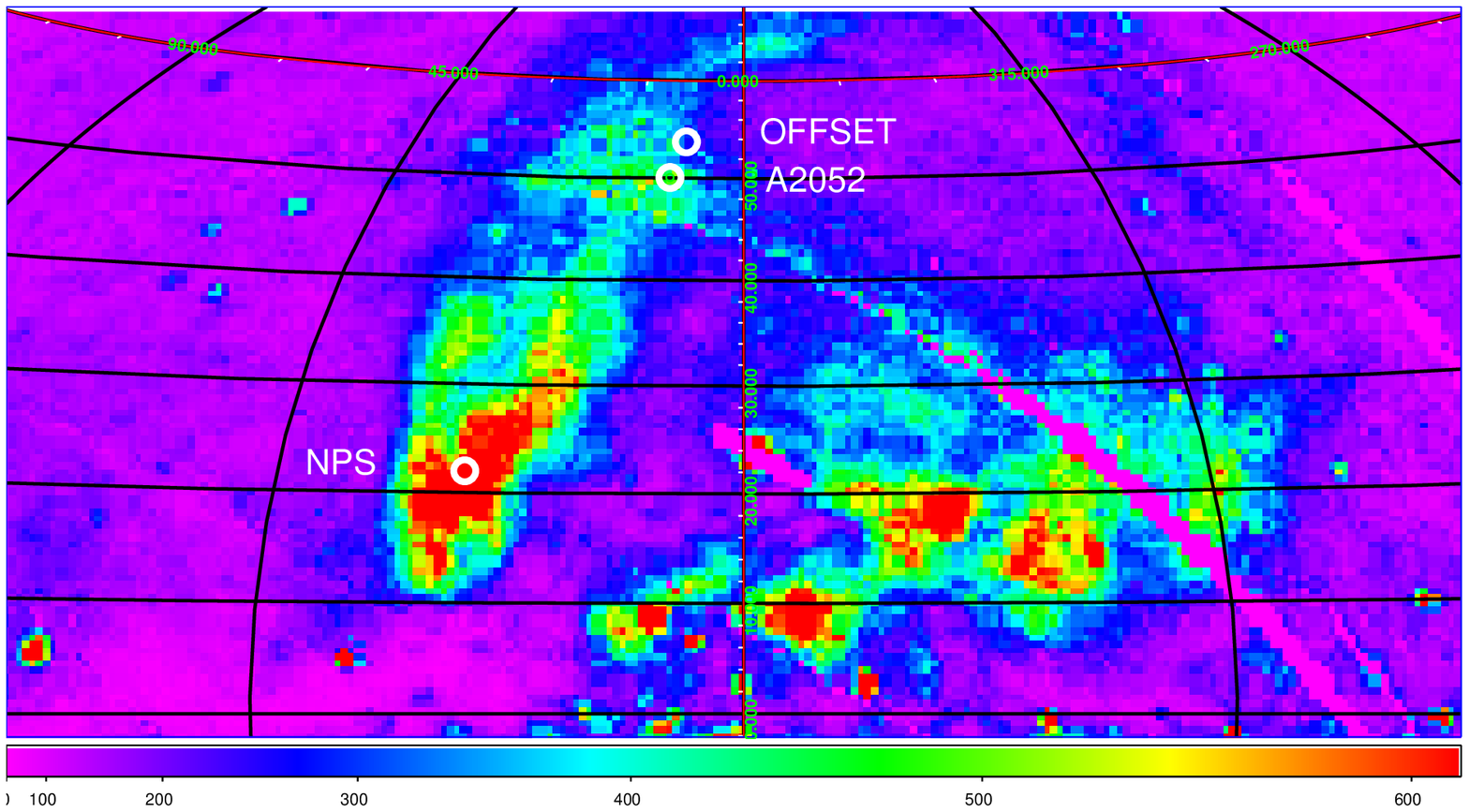,width=80mm}
\leavevmode\psfig{figure=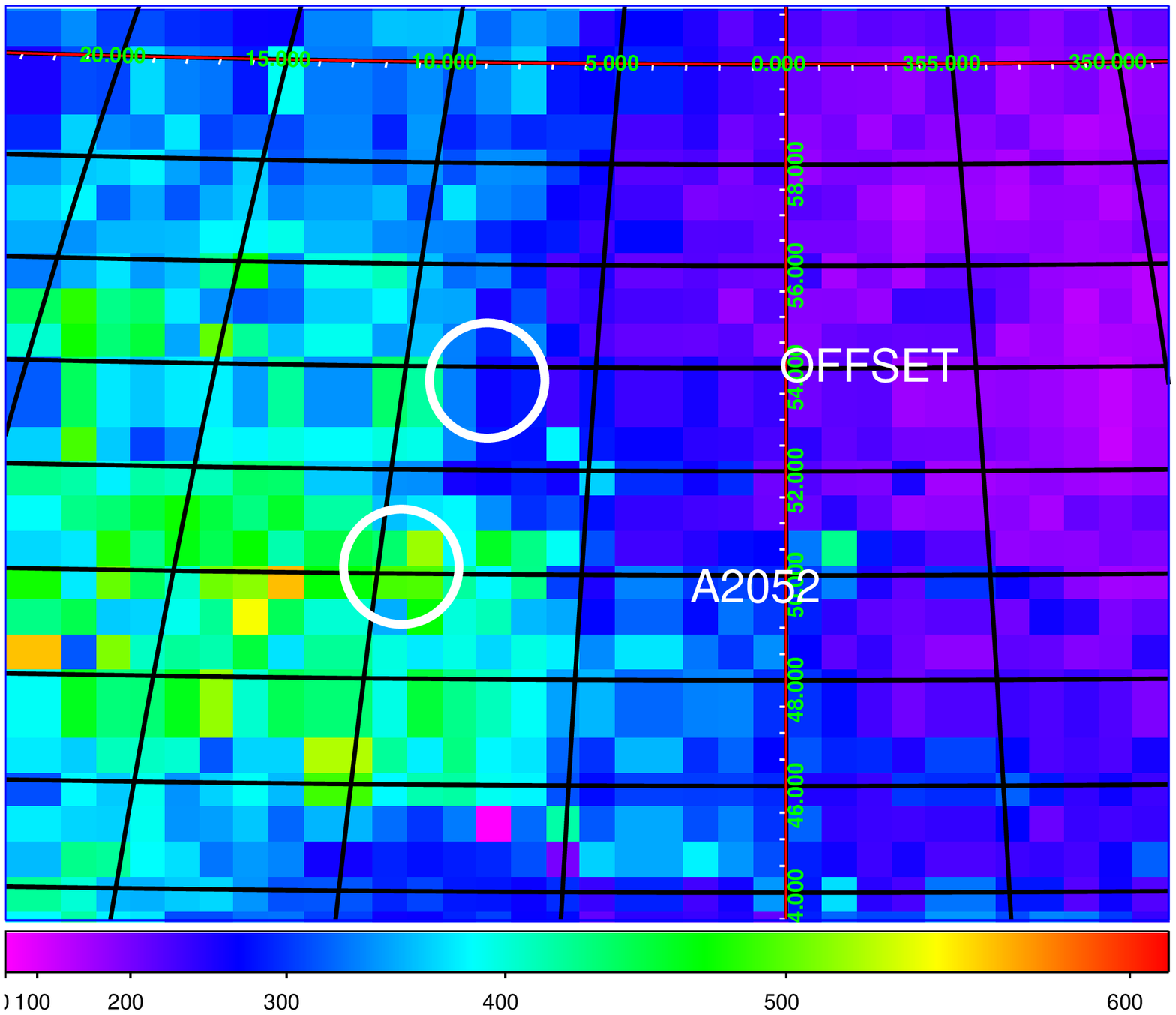,width=40mm}
  \end{center}
  \caption{{\it Suzaku} pointing positions of the A~2052, offset, and NPS are indicated on 
{\it ROSAT} 3/4~keV map (Snowden et al. 1995) in galactic coordinates.
Right panel shows a zoom-up of a part of the left panel.
The pixel size and unit are $40'.5\times 40'.5$ and cts~s$^{-1}$, respectively.
The map is obtained by NASA/HEASARC SkyView service.
}
\label{fig:rass1}
\end{figure}

Figure ~\ref{fig:rass1} shows locations of the cluster and offset pointings
overlaid on a diffuse soft X-ray map.
The bright structure extending from the bottom to the top of the map is
the North Polar Spur (NPS),  
a possible past supernova remnant or associated with local stellar winds (Egger \& Aschenbach 1995 and references therein).
We note that the count-rate in the {\it ROSAT} R34 band (3/4~keV) around the cluster
is enhanced compared to that in the offset position by 30--60\%.
Whether this is associated with the cluster, with the NPS, or with other structures
is not clear in the {\it ROSAT} map and a question to be studied in this paper.

In all observations the XIS was in the normal clocking and $3\times3$ editing modes.
While in cluster observations detectors were operated {\it without} spaced-row charge injection,
in the offset observation those were operated {\it with} spaced-row charge injection.
In this paper, we utilized only the XIS data.
Detailed description of the {\it Suzaku} observatory, the XIS instrument, 
and the X-ray telescope  are found in 
Mitsuda et al. (2007), Koyama et al. (2007), and Serlemitsos et al. (2007) respectively.

\begin{table*}[h]
  \caption{{\it Suzaku} observations of A~2052. 
}
\label{tbl:obs}
  \begin{center}
    \begin{tabular}{llllr}
\hline \hline
Name & Date & Sequence-    & (RA, Dec) & Net \\
     &      & NO   & (degree, J2000) & Exposure (ks) \\
\hline
P1 & 2005-8-20 & 100006010 & (229.2775, 6.8843) & 13.3 \\
P2 & 2005-8-20 & 100006020 & (229.3235, 7.1130) & 22.2 \\
P3 & 2005-8-21 & 100006030 & (229.0931, 7.1589) & 13.0 \\
P4 & 2005-8-21 & 100006040 & (229.0466, 6.9285) & 9.6 \\
offset & 2007-7-14 & 802038010 & (225.6293, 8.2927) & 28.7 \\
\hline
    \end{tabular}
  \end{center}
\end{table*}

\section{Analysis and Results}
\subsection{Data Reduction}
We used version 2.1 processing data along with the HEASOFT version 6.4.
We have screened event data using the standard selection criteria; i.e., 
geomagnetic cut-off-rigidity $>$ 6~GV, the elevation angle above the sunlit earth $>$\timeform{20D} and the dark earth $>$\timeform{5D}.
The regions illuminated by calibration sources at sensor edges were removed.

We examined the light curve in the 0.2--2~keV range for stable-background periods.
To obtain a good background signal, here we exclude the cluster emission within $12'$ from the X-ray center.
We found that the first $\sim6$~ks of the P4 pointing shows a count rate significantly higher than other periods. 
We also noticed that the solar-wind proton flux also shows an enhancement in the corresponding period.
The proton flux becomes as high as that in the flaring period when 
solar-wind charge-exchange X-ray emission was detected with {\it Suzaku} during the NEP pointing (Fujimoto et al. 2007).
Therefore, the high count rate in the X-ray background 
is probably caused by the charge-exchange X-ray emission.
We removed this period from the event.
The useful exposure times are given in Table.~\ref{tbl:obs}.
There is a strong X-ray source in the outskirt of the cluster at the position of (RA, Dec) $=$ (\timeform{228D.9948}, \timeform{7D.1726}). 
We removed events within $2'$ from this source.

The offset pointing data are reduced in a similar way.
We excluded events around four X-ray sources detected in this pointing.

\subsection{Spectral Fitting Method}\label{sect:method}
Combining the four pointing data, 
we extracted projected spectra in annuli around the emission center at the position of (RA, Dec) $=$ (\timeform{229D.1818}, \timeform{7D.0348}). 
The inner and outer radii are $3', 6', 9', 12', 15'$, and $20'$ (Fig.~\ref{fig:xisimage}).
Note that the outermost extraction covers only a part of the annulus.
The central region ($r<3'$) is excluded to avoid complexity related with the central cool component of the cluster.
Here and hereafter we denote $r$ as a projected radius from the cluster center.
We also examine the spectrum integrated from $9'<r<20'$ for properties averaged over the outer part.
The three front-illuminated (FI) CCD data are co-added together.
Then the FI and back-illuminated (BI) CCD spectra are fitted simultaneously with a common model but with different normalizations.

We subtracted the instrumental (Non-X-ray) background following a method in Tawa et al. (2008) using the software xisnxbgen.
In this method, the background spectrum is generated by summing up the dark earth data sorted 
with the same fractional distribution of the HXD/PIN UD count rates and from the same extraction as the source.
This method provides an accurate reproduction of the background
with a systematic uncertainty of less than 10\% (Tawa et al 2008).

The lowest energies to use from the FI and BI CCDs are
0.4~keV and 0.32~keV, respectively.
The highest energies are 4.0--7.1~keV depending on the signal to background ratio.
To avoid calibration uncertainties around the Si edge 
we remove the energy range of 1.825--1.840~keV.
We prepared an X-ray telescope response function for each spectrum using the XIS ARF builder software xissimarfgen (Ishisaki et al. 2007).
The accumulating contamination of the XIS is also taken into account for responses using the calibration file version 2006-10-16.
Because cluster observations were done just 7-8 days after the XIS door-open,
the contamination was not large.
The reduction of the efficiency is estimated to be 8~\% at maximum.
The contamination for the offset pointing data was much larger.
To made the XIS response function, we used the software xisrmfgen (version 2007-05-14).
These responses are made assuming an uniform brightness distribution of the source, 
since our main target, the soft excess emission, is diffuse and distributed over the field of view.

To describe the thermal emission from the collisional ionization equilibrium (CIE) plasma, 
we use the APEC model (Smith \& Brickhouse 2001) with the solar metal abundances
taken from Anders \& Grevesse (1989), unless stated otherwise.

After some trials, we found that the cosmic X-ray background (CXB) in the hard band 
can be described well by the standard power-law model
with a reported photon index, $1.4$, and normalization of 10 ph cm$^{-2}$ s$^{-1}$ sr$^{-1}$ keV$^{-1}$.
We use this fixed model for the CXB, 
unless stated otherwise.
This component 
is assumed to be absorbed by the Galactic neutral gas.
Column densities in the cluster and offset pointings are assumed to be 
$2.72\times 10^{20}$ cm$^{-2}$ and $2.2\times 10^{20}$ cm$^{-2}$, 
respectively, based on the H\emissiontype{I} map \citep{dickey90}.
We use the photo-electric absorption of Wisconsin cross-sections (wabs model in the XSPEC).  

\subsection{The Offset-Pointing Spectra} \label{sect:offset}
\begin{figure}[htbp]
\begin{center}
\leavevmode\psfig{figure=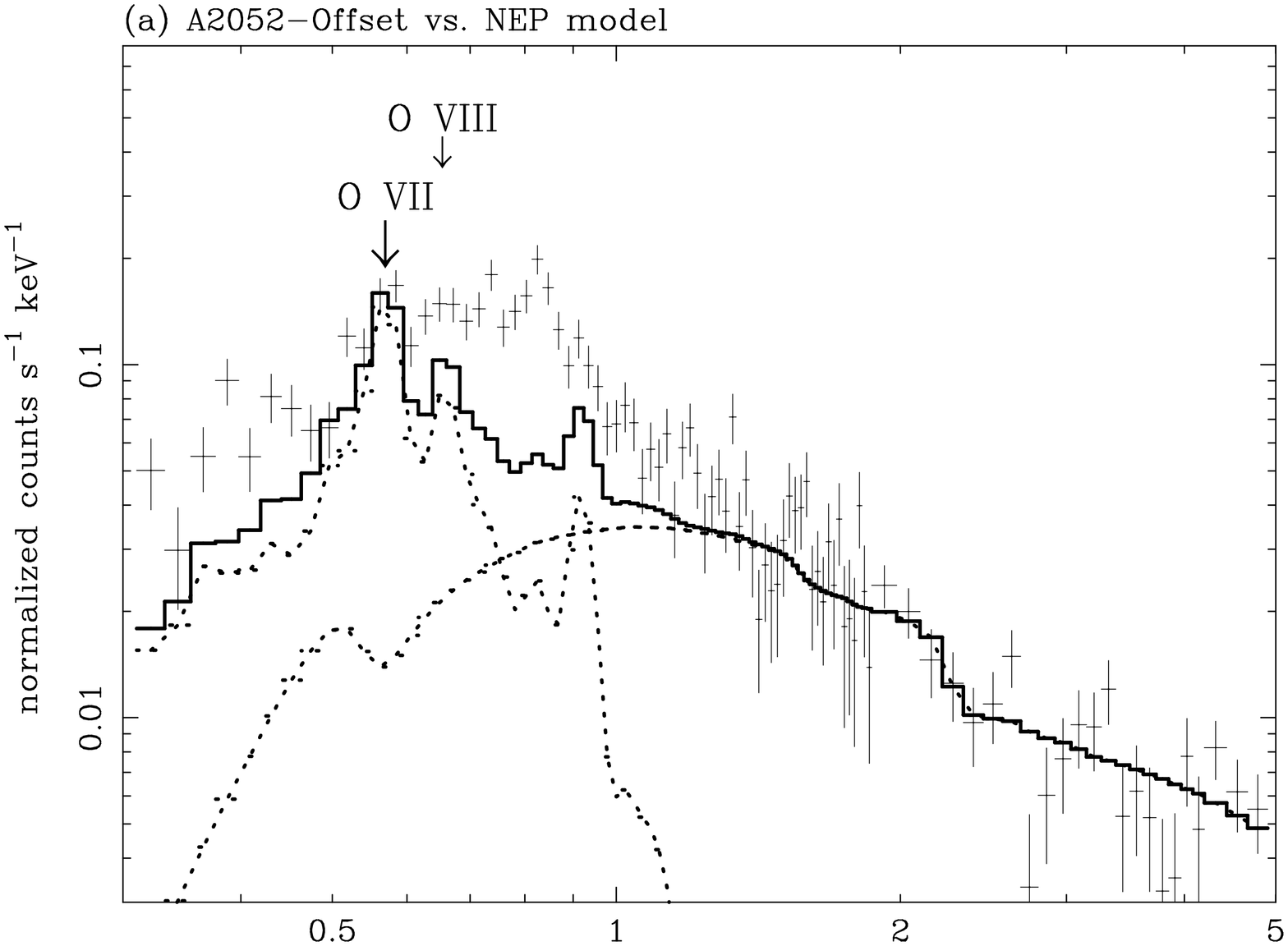,angle=-90,width=0.60\textwidth}
\leavevmode\psfig{figure=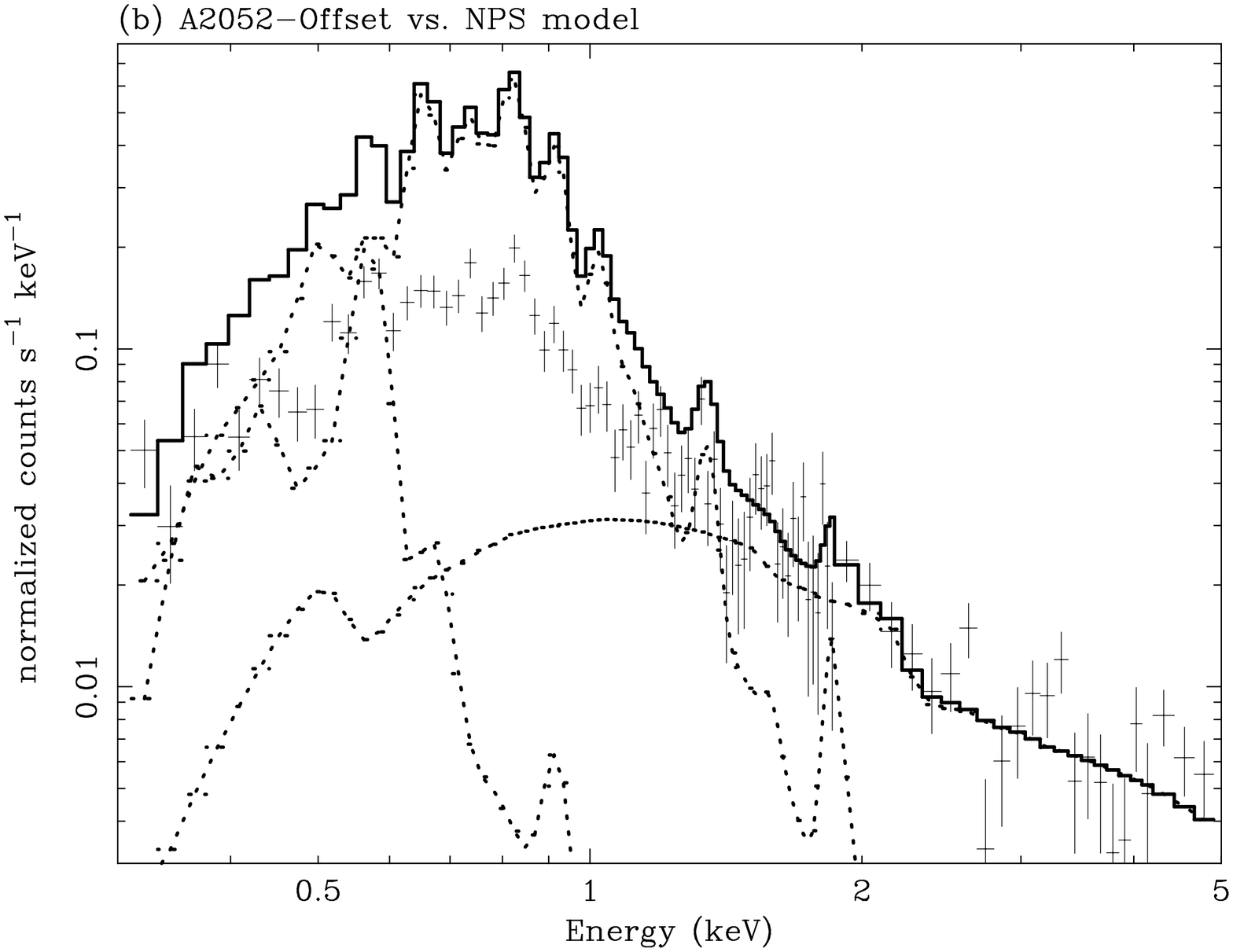,angle=-90,width=0.60\textwidth}
\end{center}
  \caption{(a) A~2052 offset spectrum (data points) is compared with a spectral model from another blank sky field, NEP (histogram).
The spectral model and its components (dotted-lines) are taken from Fujimoto et al. (2007). The positions of oxygen energies are indicated.
(b) Same as (a), but the model is taken from the brightest point of the NPS region (Miller et al. 2008).
}
\label{fig:vsNEP}
\end{figure}

In order to estimate the Galactic foreground emission,
we analyzed the spectrum from the offset pointing.
Here we compare the A~2052 offset spectrum with that from another blank-sky,
the NEP region.
In Fig.~\ref{fig:vsNEP}(a), we show the offset spectrum along with the best-fit model from the NEP.
The NEP model is based on the XIS data in the period when the sky background is stable (Fujimoto et al. 2007).
The Galactic coordinates ($l,b$) of the A~2052 offset and NEP are (\timeform{9D.4}, \timeform{50D.1}) and 
(\timeform{95D.8}, \timeform{25D.7}), respectively. 
The two O\emissiontype{VII} line (rest frame energy of 569~eV) fluxes are similar.
The O\emissiontype{VII} line flux from the offset is derived to be 7.1~photons cm$^{-2}$ s$^{-1}$ sr$^{-1}$ (or line units, LU).
This flux is comparable to those from other blank sky positions with {\it Suzaku}, 
including the MBM~12 (molecular cloud) off-cloud without subtracting on-cloud flux (5.9~LU; Smith et al. 2007) and
the A~2218 offset-A pointing ($\sim 8$~LU; Takei et al. 2007).
On the contrary, 
below and above the oxygen line up to $\sim 1.4$~keV, 
the A~2052 offset spectrum shows clear enhancements
compared with the NEP model.
Note that these residuals can be seen neither in the NEP data.
These variations not only in the flux but also in the spectral shape 
require a careful estimation of the foreground emission. 

In the offset spectrum,
there are emission lines at the O\emissiontype{VII} ($T_{\rm max} = 0.18$~keV) and 
O\emissiontype{VIII} (654~eV; $T_{\rm max} = 0.26$~keV) positions.
Here $T_{\rm max}$ is the temperature where the emissivity at the CIE condition becomes maximum.
In addition, there are line structures around $\sim 0.7$~keV and $\sim 0.8$~keV.  
These could be due to a combination 
of Fe\emissiontype{VII} ($T_{\rm max} = 0.6$~keV), 
Fe\emissiontype{VIII} ($T_{\rm max} = 0.6$~keV), and other Fe-L lines.
These oxygen and iron line structures are difficult to form from a single temperature CIE plasma with the solar metal abundance ratio.

\begin{figure}[htbp]
\begin{center}
\leavevmode\psfig{figure=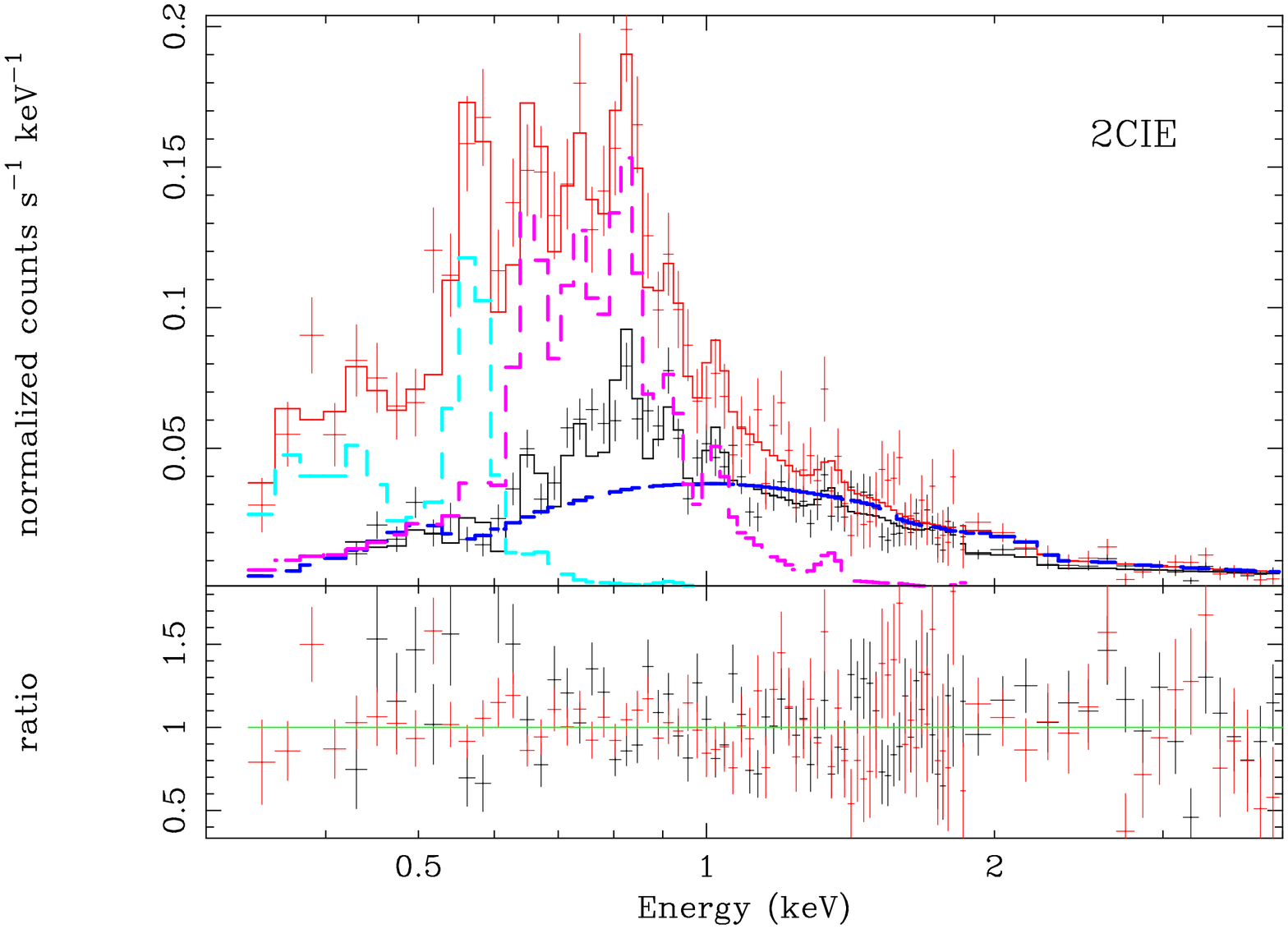,angle=-90,width=0.6\textwidth}
\leavevmode\psfig{figure=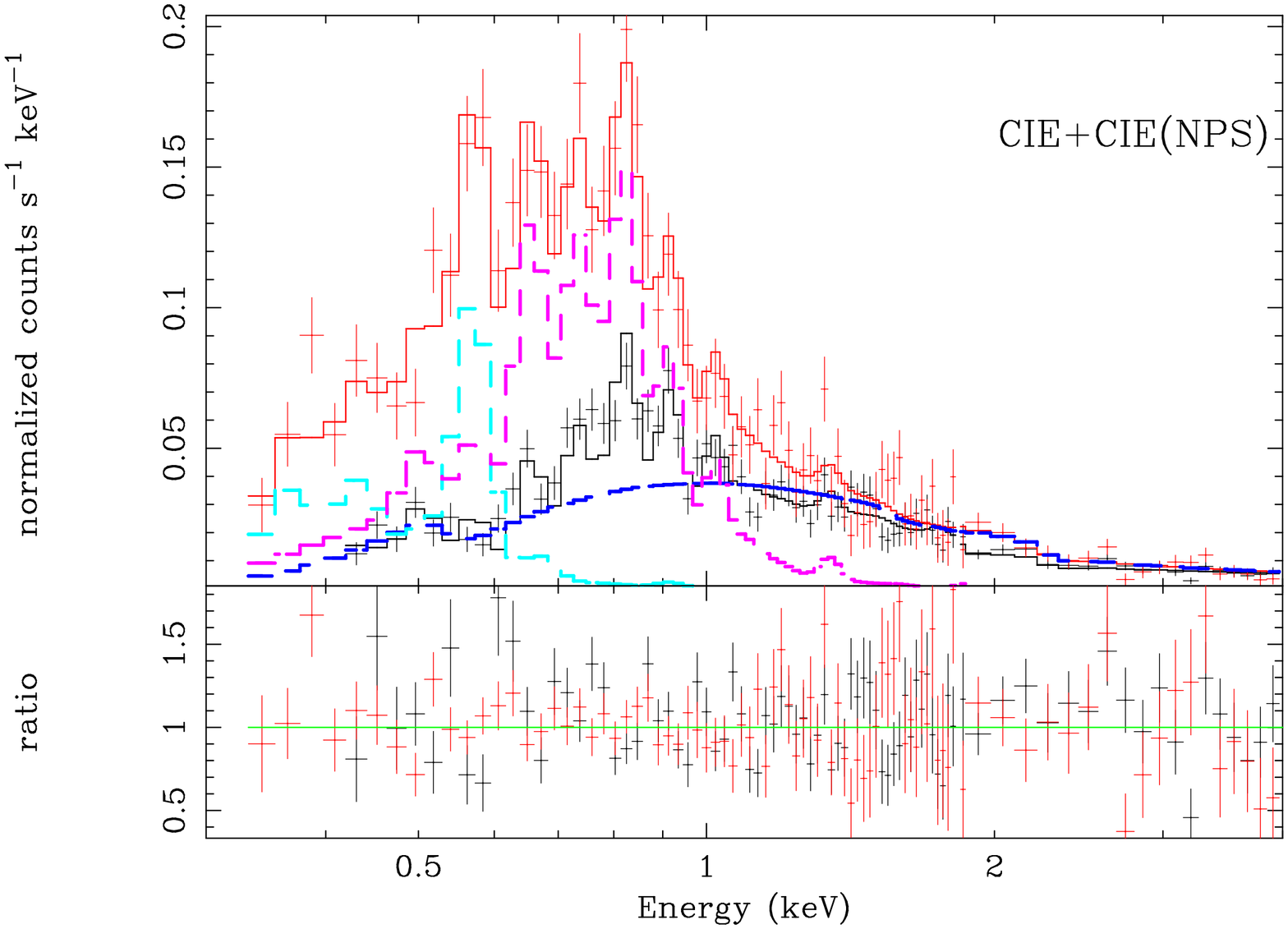,angle=-90,width=0.6\textwidth}
\end{center}
  \caption{Spectral fitting result from the offset pointing data.
Upper panel: The FI and BI CCD data are shown in black and red colors, respectively.
Solid lines show the best-fit model (2 CIE plus CXB).
Model components only for the BI CCD are shown in dashed lines.
Lower panel: Fit residuals in terms of the data to model ratio.
}
\label{fig:spe-bgd}
\end{figure}

We found that the spectra can be described by a combination of two CIE emission components,
as shown in Fig.~\ref{fig:spe-bgd} and Table~\ref{tbl:fit-bgd}.
At energies below 0.4~keV, the fit is not good.
To avoid a possible effect of this residual on the Galactic foreground modeling,
we ignore this energy band here and hereafter.
New best-fit parameters are given in Table~\ref{tbl:fit-bgd} and are similar to those of the previous fit.
We confirmed that the observed flux is consistent with the count in this direction in the ROSAT All-Sky Survey diffuse background map.

\begin{table*}[h]
  \caption{The fitting results for the offset pointing spectrum.
}
\label{tbl:fit-bgd}
  \begin{center}
    \begin{tabular}{llllllll}
\hline \hline
Model & $\Delta E$ & $\chi^2$ & d.o.f.& $kT_{\rm 1}$ & Norm$_1$\footnotemark[$*$] & $kT_{\rm 2}$ & Norm$_2$\footnotemark[$*$] \\
      & (keV) &          &       &  (keV) & & (keV) & \\
\hline
2CIE   & 0.32--4.0 & 236      & 165   & 0.096$\pm 0.01$ & $5.2\pm 0.6$ & $0.34\pm 0.02$& $1.5\pm0.1$ \\
2CIE & 0.4--4.0 & 229      & 162   & 0.094$\pm 0.01$ & $6.7\pm0.8$ & $0.34\pm 0.02$& $1.5\pm0.1$ \\
CIE+CIE(NPS) & 0.4--4.0   & 220      & 162   & 0.095$\pm 0.01$  & $5.2\pm 0.6$ & $0.31\pm0.01$ & $3.8\pm 0.2$ \\
\hline
\multicolumn{8}{l}{ \parbox{150mm}{\footnotesize
\footnotemark[$*$] 
$10^{-17}(4\pi)^{-1} D_\mathrm{A}^{-2}(1+z)^{-2}\int n_\mathrm{e}n_\mathrm{H}~dV$, 
where $n_\mathrm{e}$ and $n_\mathrm{H}$ are the electron and hydrogen density (cm$^{-3}$) 
and $D_\mathrm{A}$ is the angular size distance to the source (cm).
This is scaled for a sky area of 1257 arcmin$^2$ ($20'$ radius circle).
}
}
    \end{tabular}
  \end{center}
\end{table*}

As seen in Fig.~\ref{fig:rass1} presented in \S~\ref{sect:obs}, 
the offset and A~2052 positions are at a tips of the NPS.
This figure suggests that the foreground emission 
in these directions could be contaminated by that associated with the NPS. 
In fact, a large part of the hotter component of the offset is most likely associated with the NPS emission, as discussed in the next section.
For comparison, we give the best-fit model for the {\it Suzaku} spectrum 
of the brightest position of the NPS (Miller et al. 2008)
in Fig.~\ref{fig:vsNEP}(b) along with the offset spectrum.

We also attempt to use the emission model
derived from the {\it Suzaku} observation of the NPS (Miller et al. 2008).
Here we use different abundance ratios instead of the solar abundance ratio used above for the hotter component.
Those ratios are as follows;
C=0.0, N=1.33, O=0.33, Ne=0.51, Mg=0.46, Fe=0.5, relative to the solar.
This change improved the fit slightly [$\Delta \chi^2 = 9$; Table~\ref{tbl:fit-bgd}; CIE+CIE(NPS) model].
An additional absorption does not improve the fit significantly.
For the simplicity sake, we use the 2CIE model (solar abundance ratio; 0.4--4.0~keV) as the Galactic foreground below unless stated otherwise.

\subsection{The Soft Excess Emission} \label{sect:cluster-spectra}
We investigate spectral properties of the soft excess emission 
in the cluster direction.
First, we assume that the observed spectra consist of
the Galactic foreground, ICM, CXB, and instrumental background components.
We fixed the Galactic emission parameters as derived from the offset pointing above (\S~\ref{sect:offset}; 2CIE model).
The ICM is modeled by a single temperature CIE with a variable metal abundance.
This emission is assumed to be absorbed by the Galactic neutral gas.
The last two background components are treated as explained above (\S~\ref{sect:method}).
We refer to this as 'no excess model'.
The fitting parameters are shown in the first row of Table~\ref{tbl:fit}.
This model can not describe the data in all radial regions.
Significant residuals around the position of O\emissiontype{VII} line are seen commonly at least in $r>9'$ spectra.

\begin{figure*}
\begin{center}
\leavevmode\psfig{figure=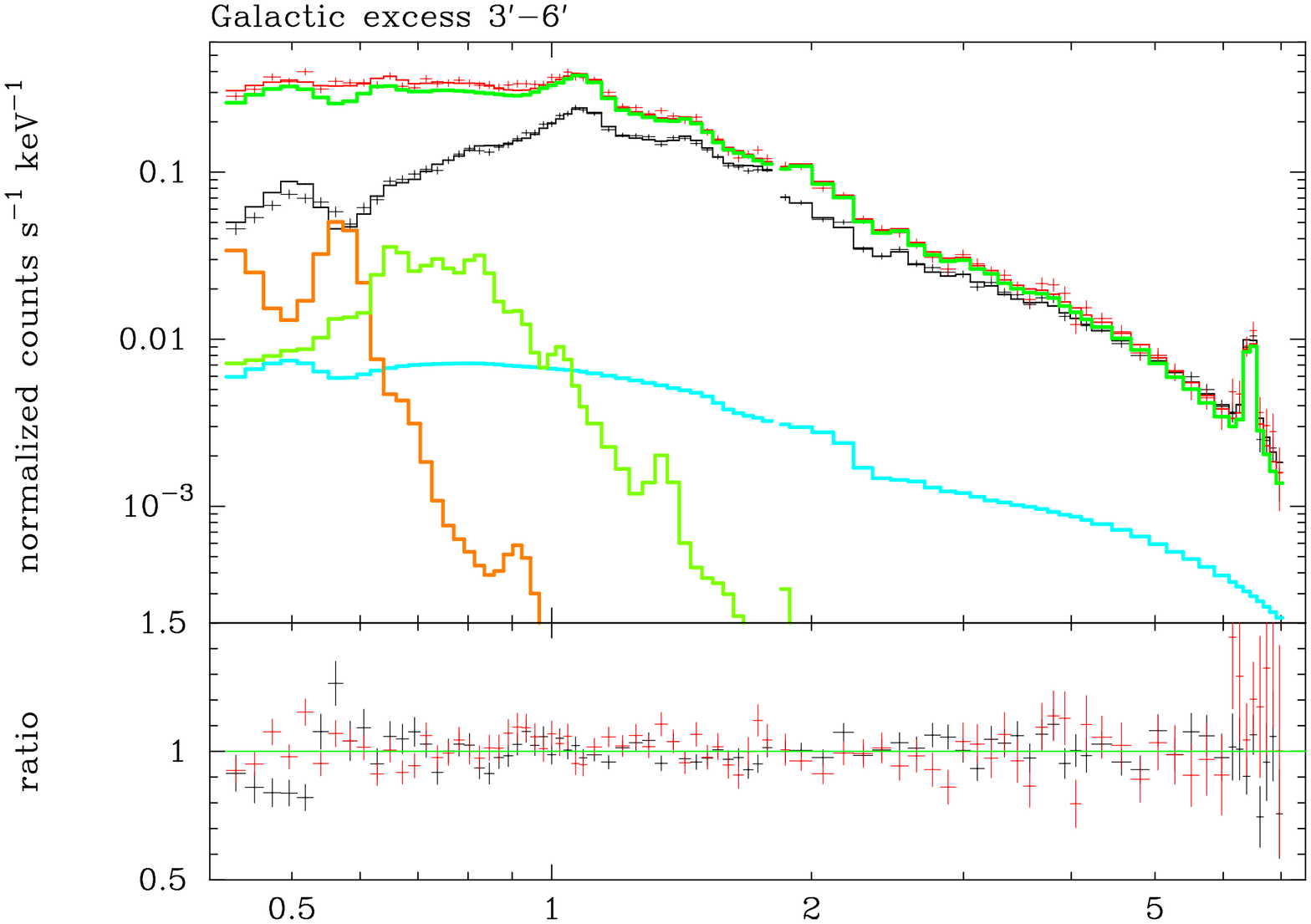,angle=-90,height=0.2\textheight,width=0.45\textwidth}
\leavevmode\psfig{figure=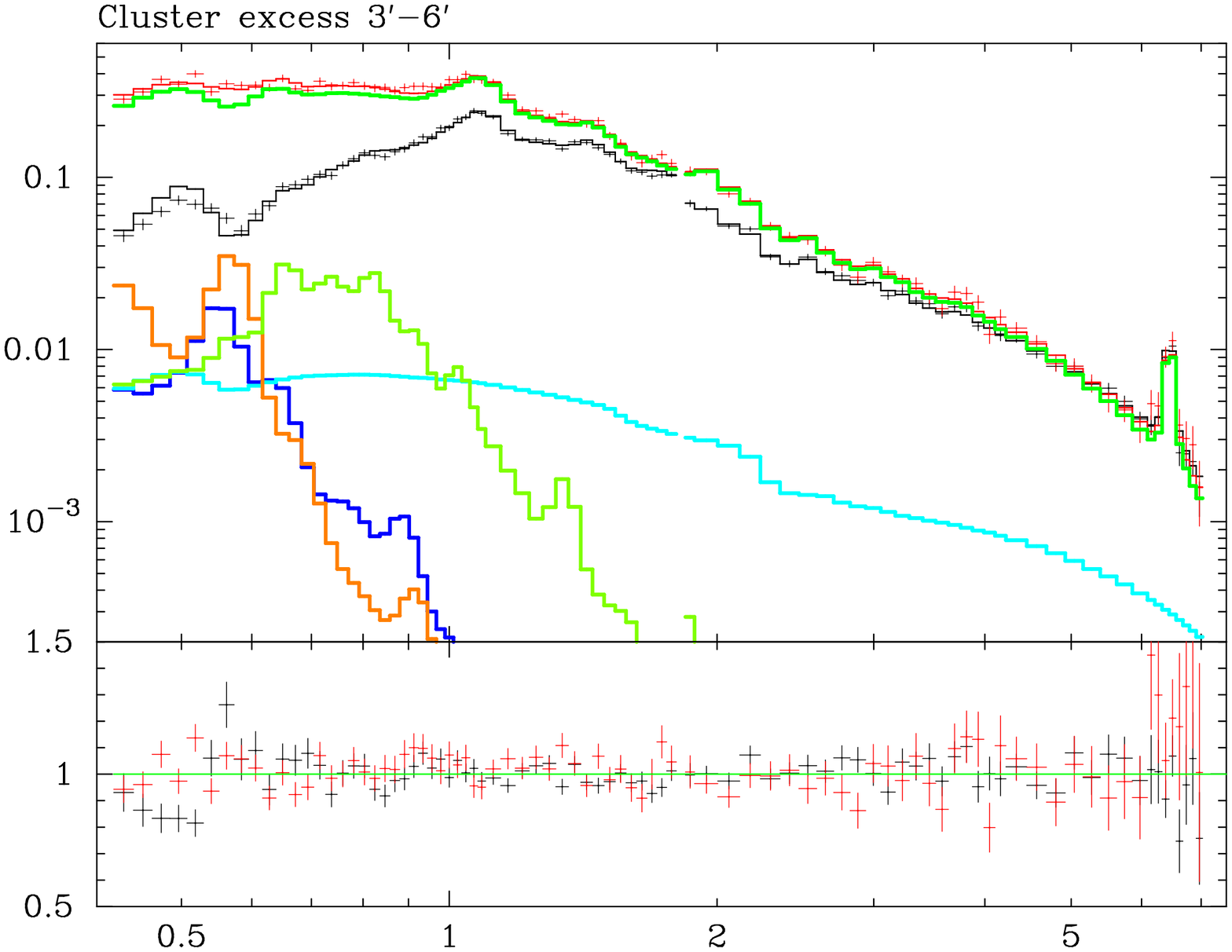,angle=-90,height=0.2\textheight,width=0.45\textwidth}
\leavevmode\psfig{figure=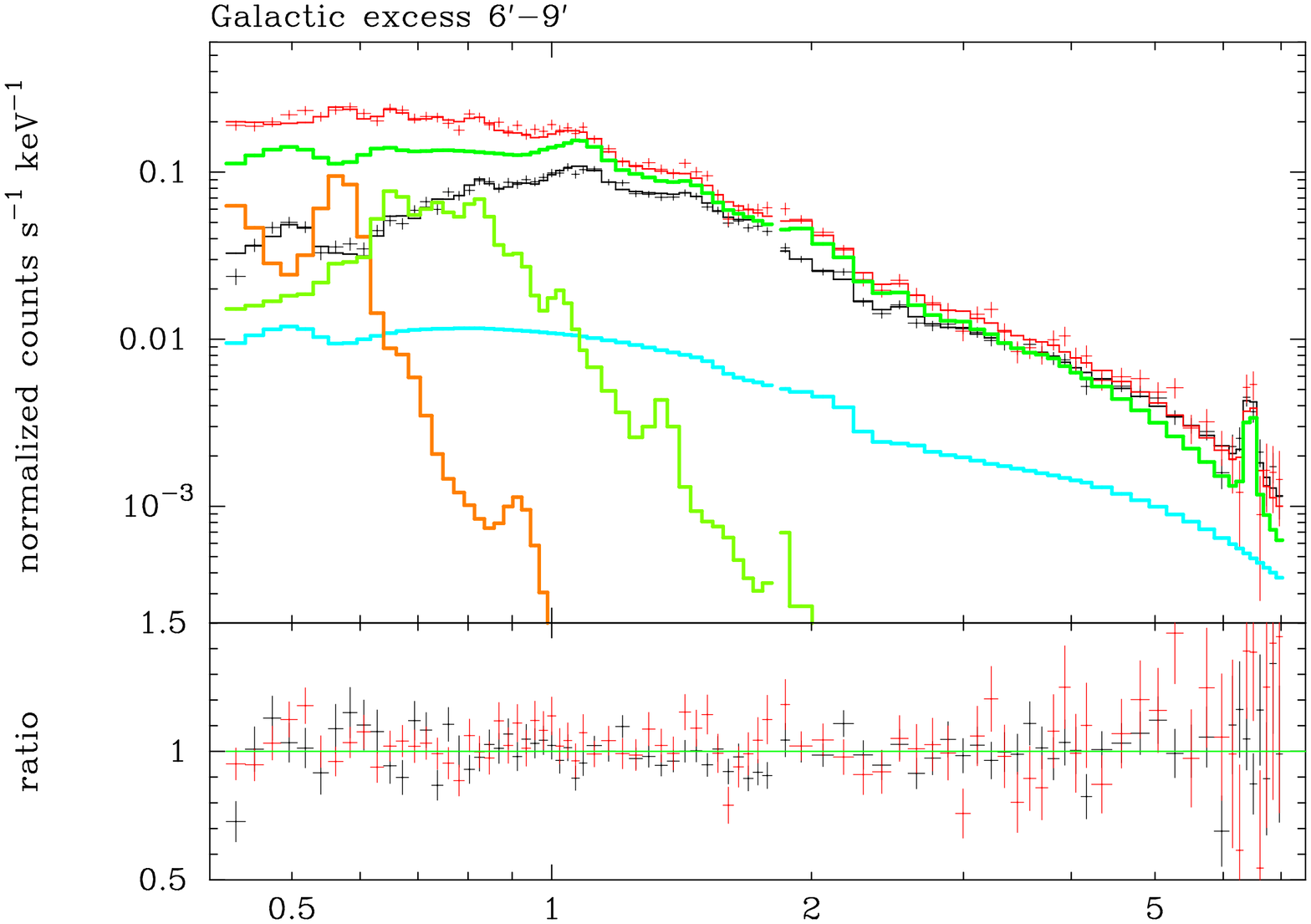,angle=-90,height=0.2\textheight,width=0.45\textwidth}
\leavevmode\psfig{figure=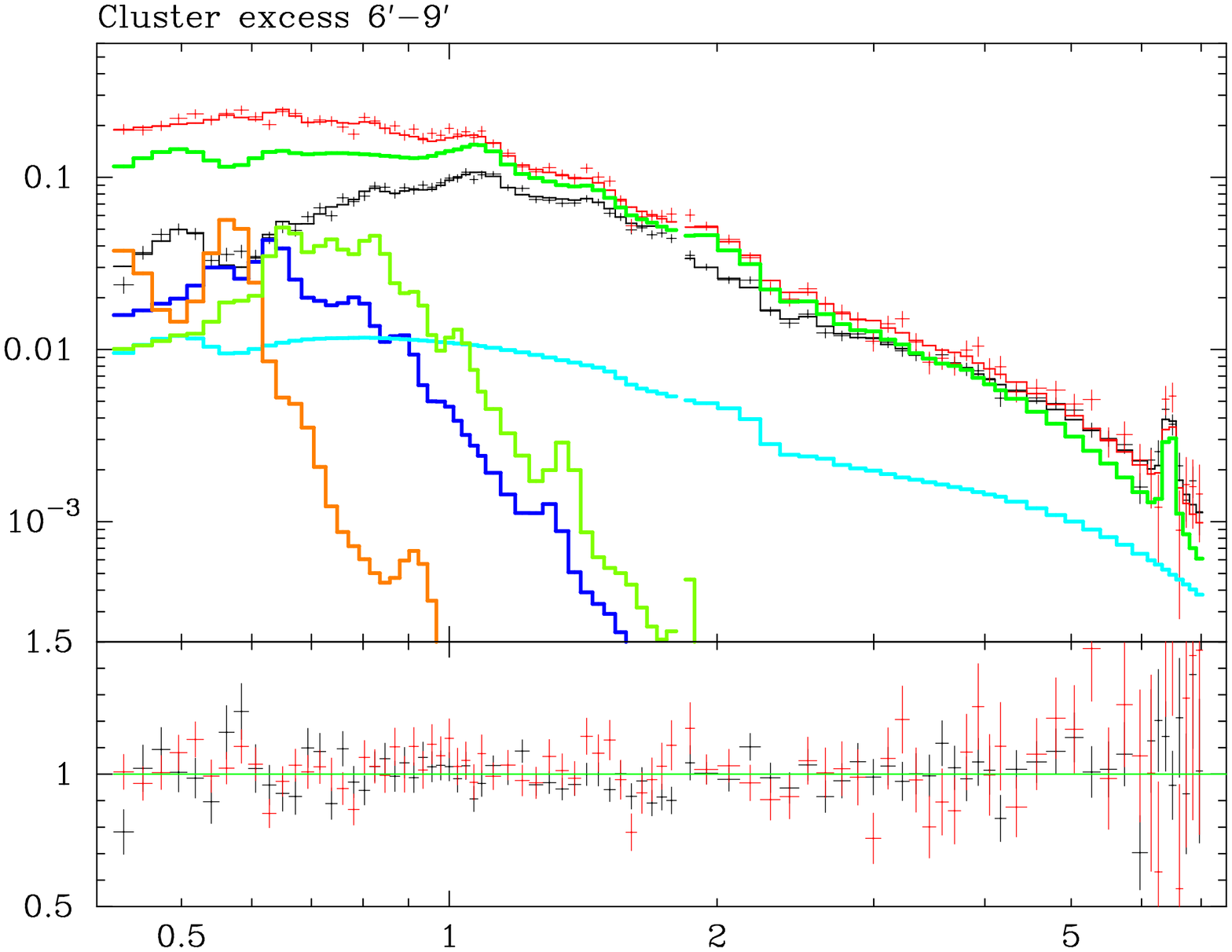,angle=-90,height=0.2\textheight,width=0.45\textwidth}
\leavevmode\psfig{figure=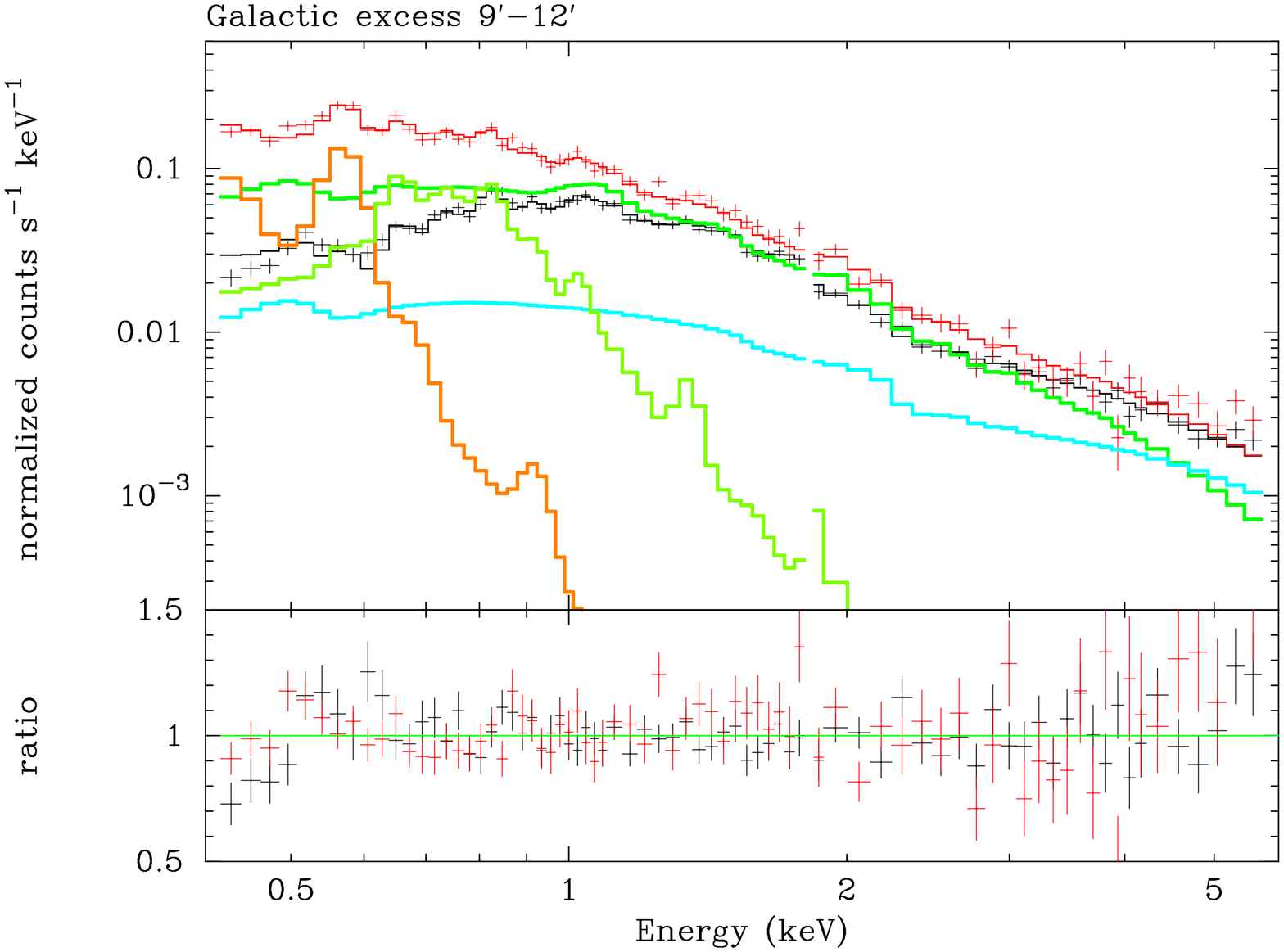,angle=-90,height=0.2\textheight,width=0.45\textwidth}
\leavevmode\psfig{figure=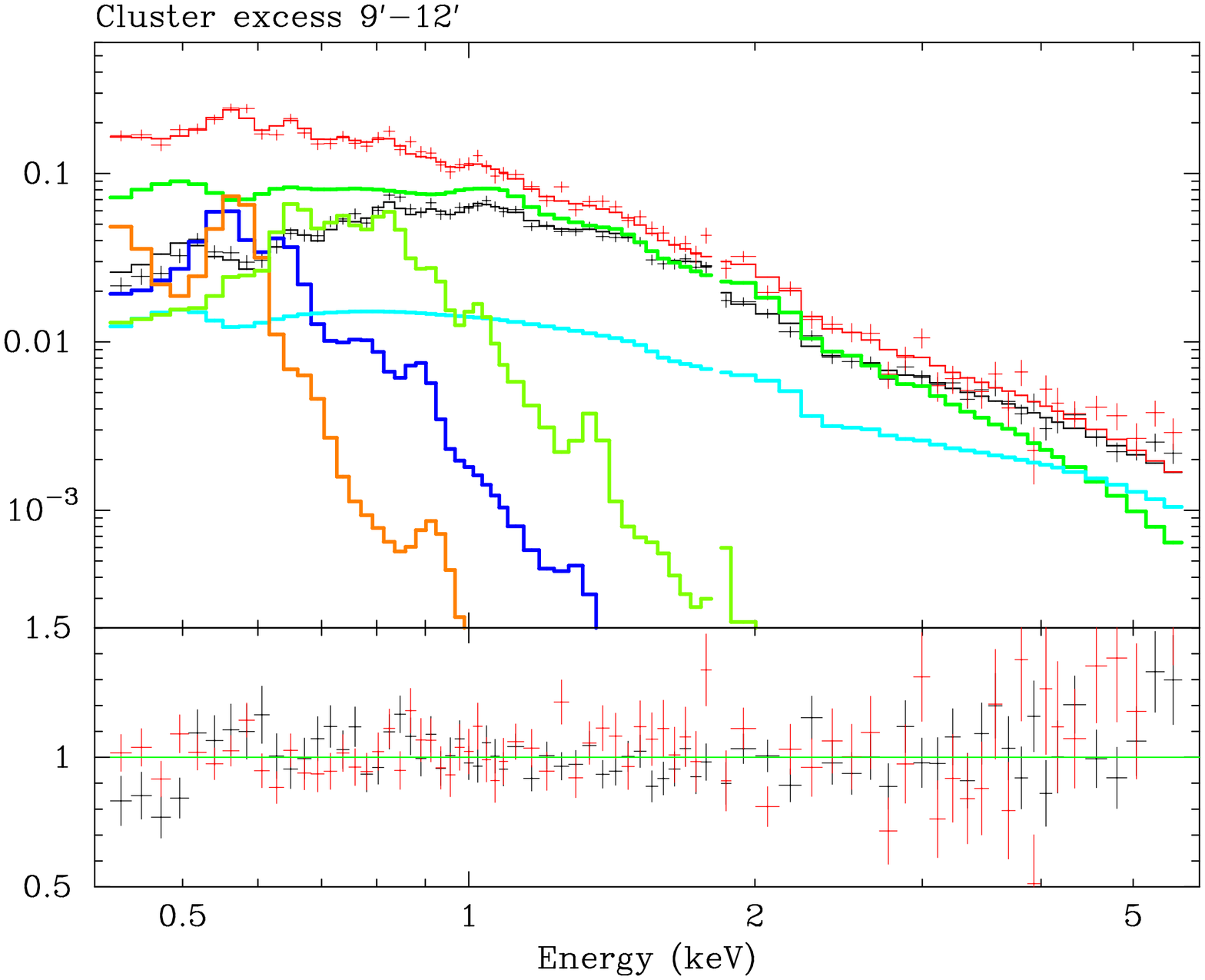,angle=-90,height=0.2\textheight,width=0.45\textwidth}
\end{center}
  \caption{Cluster spectra fitted with the Galactic excess (left panels) and cluster excess (right panels) models.
From top to bottom, the data from $3'-6'-9'-12'$ regions are shown.
The FI CCD data and BI one are shown in black and red colors, respectively.
The model components are shown by histograms with different colors.
Left panels: green, ICM; orange, 0.094~keV CIE; light-green, 0.34~keV CIE; light-blue, CXB.
Right panels: green, ICM; blue, additional CIE; orange, 0.094~keV CIE; light-green, 0.33~keV CIE; light-blue, CXB.
}
\label{fig:spec1}
\end{figure*}

\begin{figure*}
\begin{center}
\leavevmode\psfig{figure=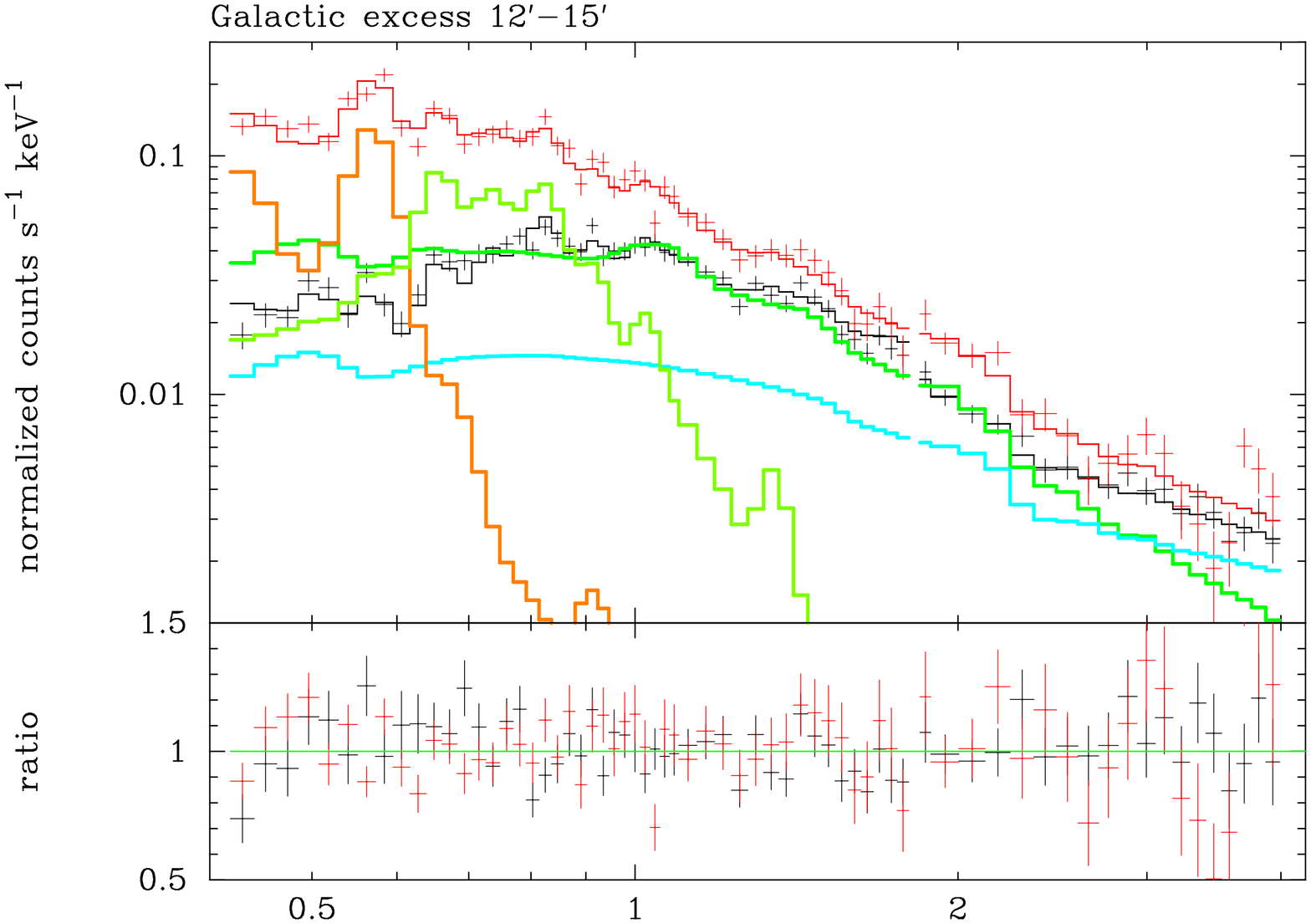,angle=-90,height=0.2\textheight,width=0.45\textwidth}
\leavevmode\psfig{figure=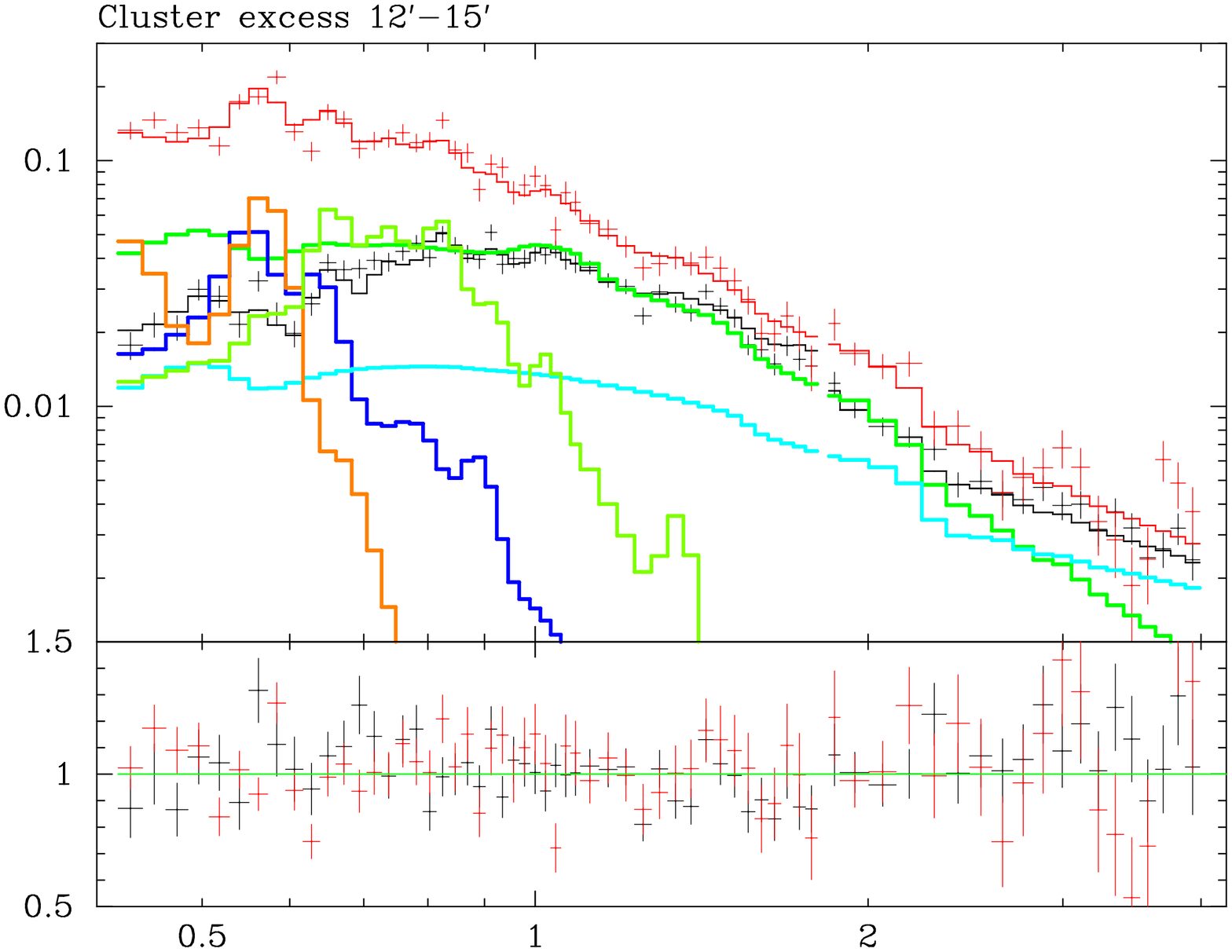,angle=-90,height=0.2\textheight,width=0.45\textwidth}
\leavevmode\psfig{figure=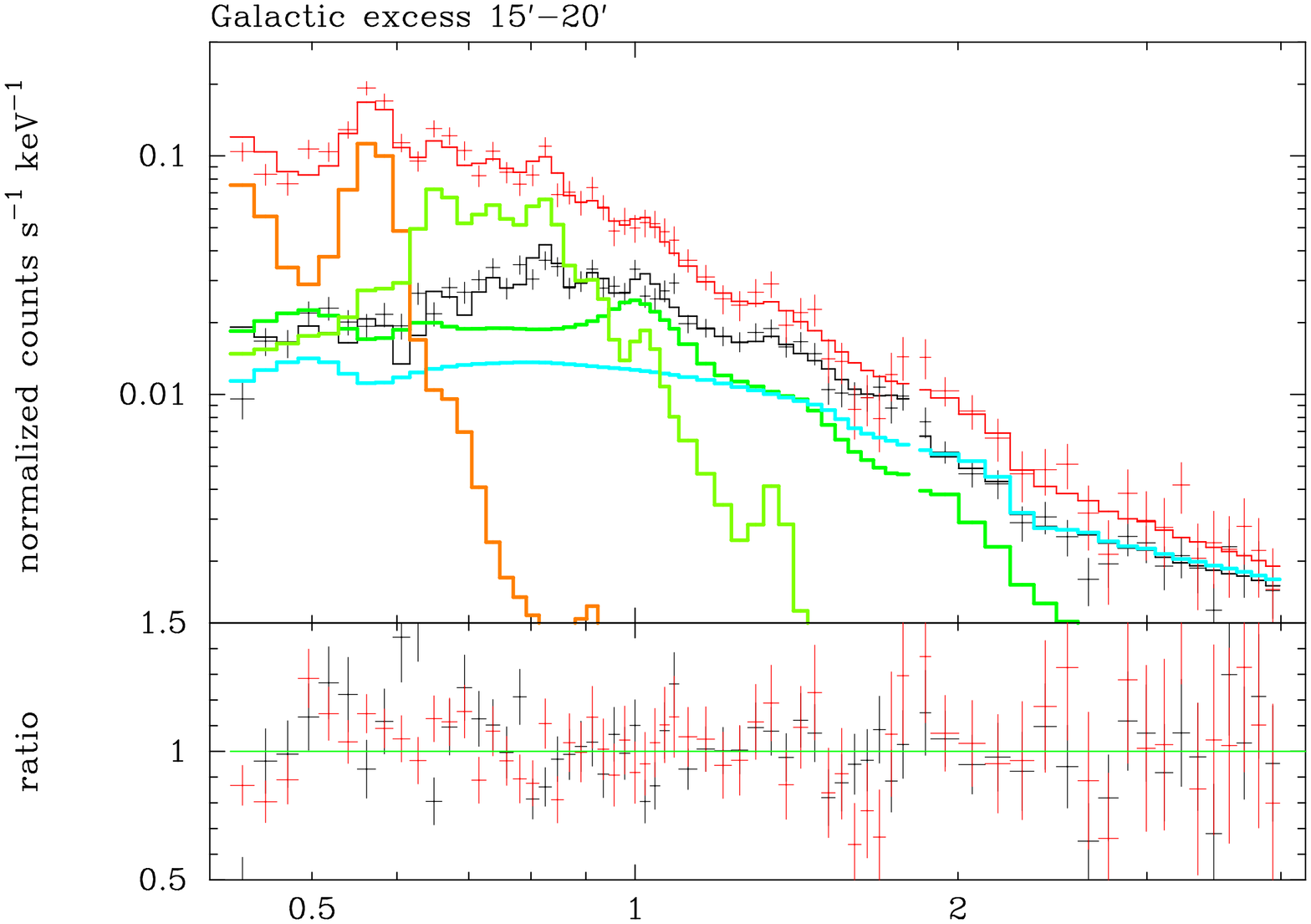,angle=-90,height=0.2\textheight,width=0.45\textwidth}
\leavevmode\psfig{figure=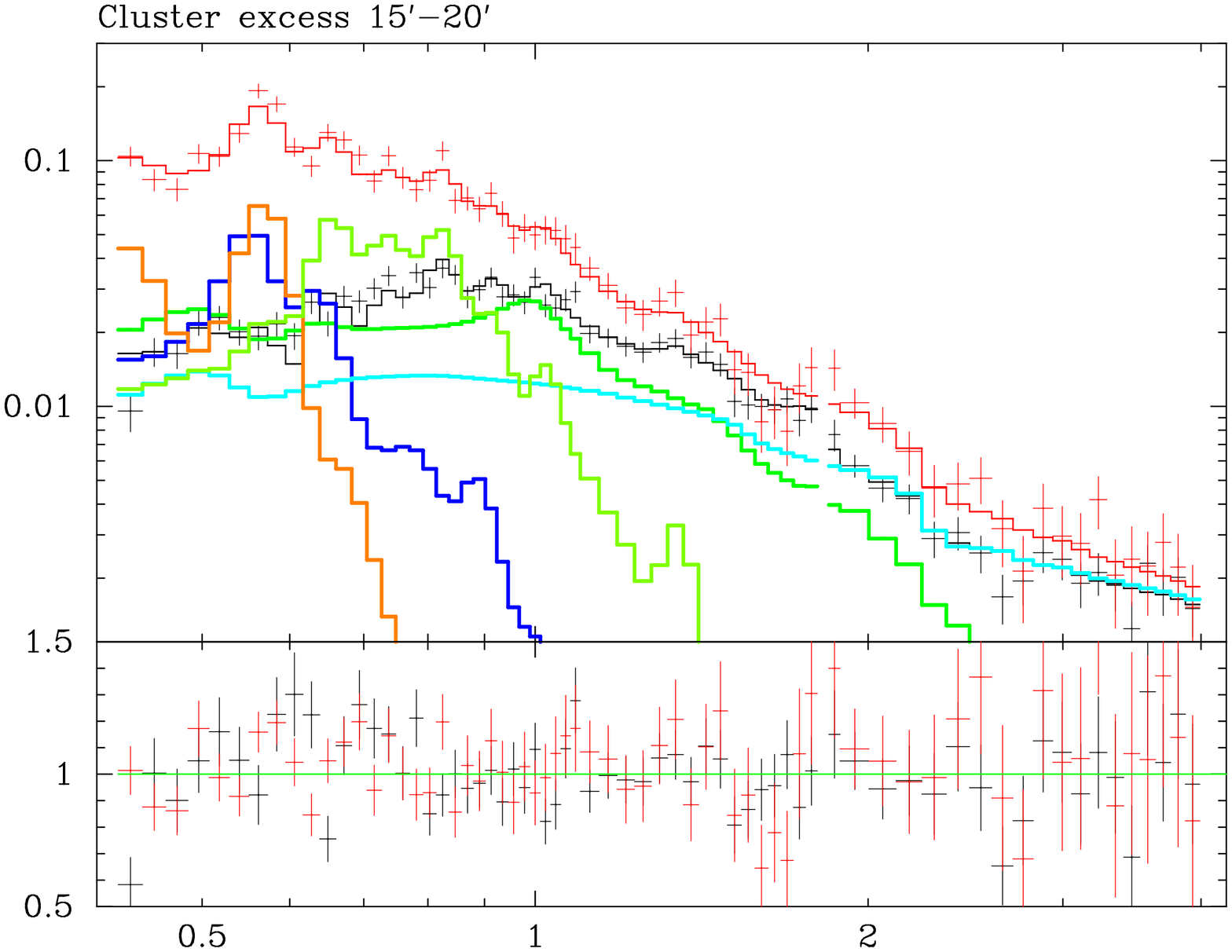,angle=-90,height=0.2\textheight,width=0.45\textwidth}
\leavevmode\psfig{figure=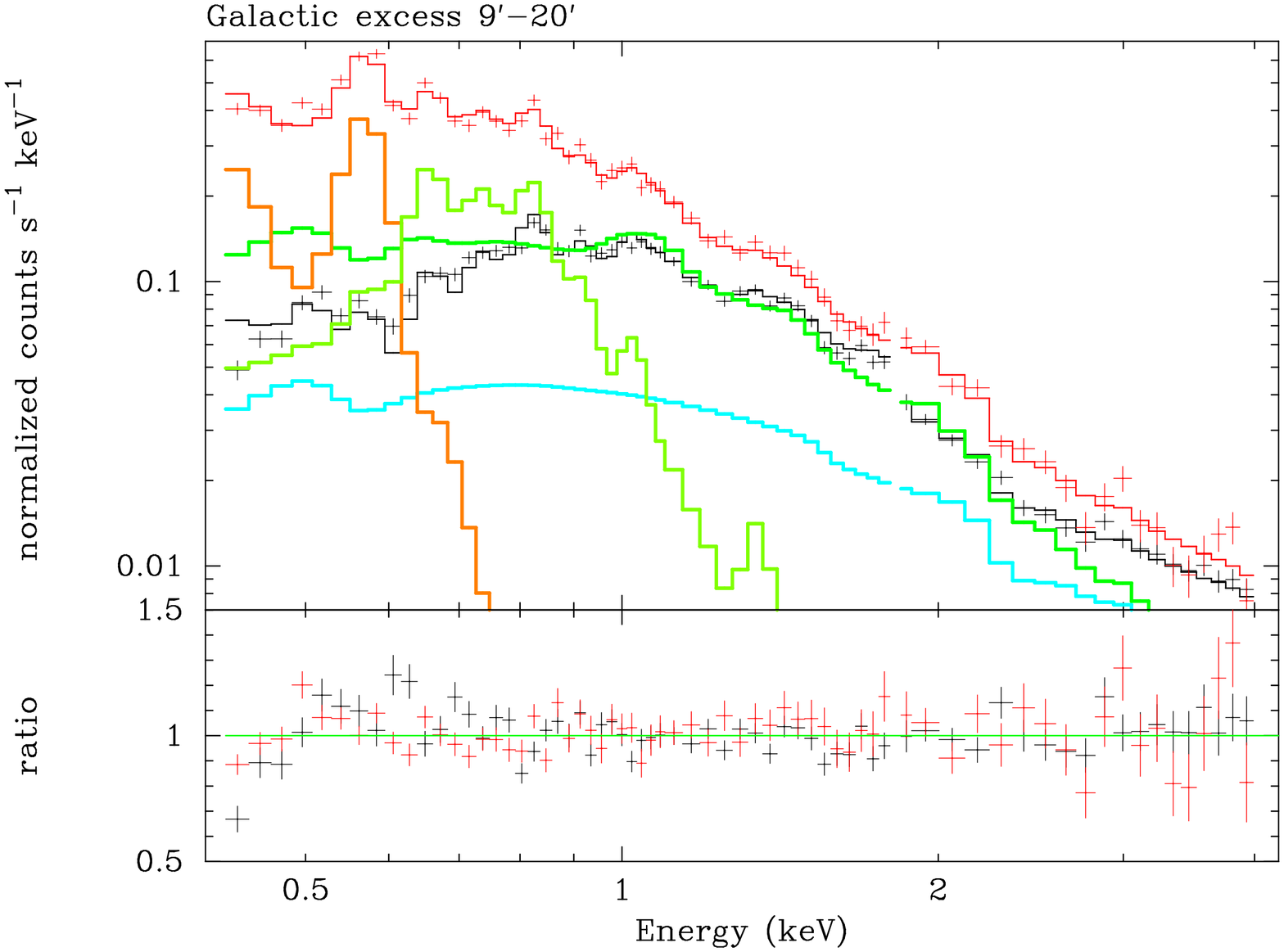,angle=-90,height=0.2\textheight,width=0.45\textwidth}
\leavevmode\psfig{figure=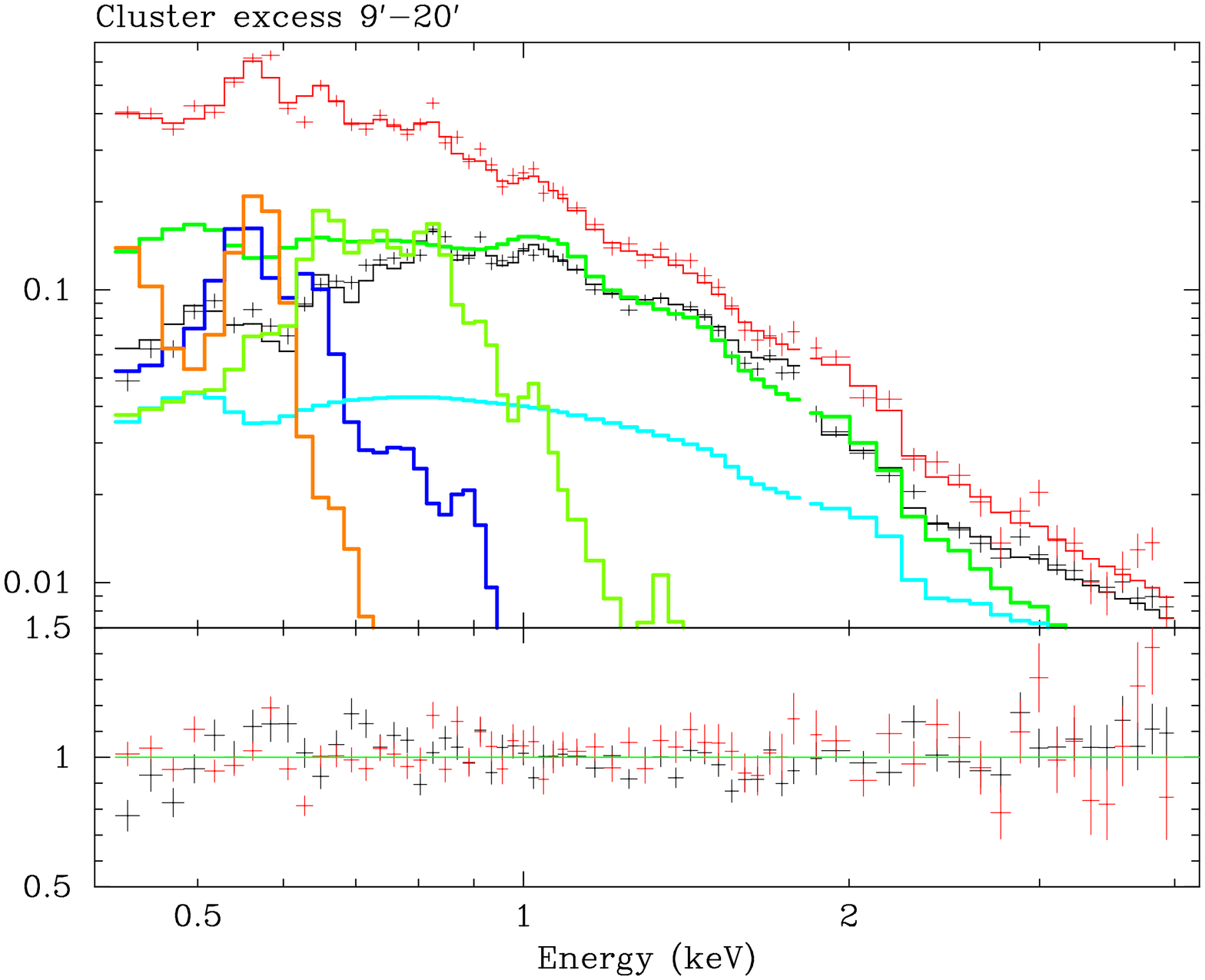,angle=-90,height=0.2\textheight,width=0.45\textwidth}
\end{center}
  \caption{Same as the previous figure, but data from $12'-15'-20'$ and $9'-20'$ regions.}
\label{fig:spec2}
\end{figure*}

\begin{figure}[htbp]
\begin{center}
\leavevmode\psfig{figure=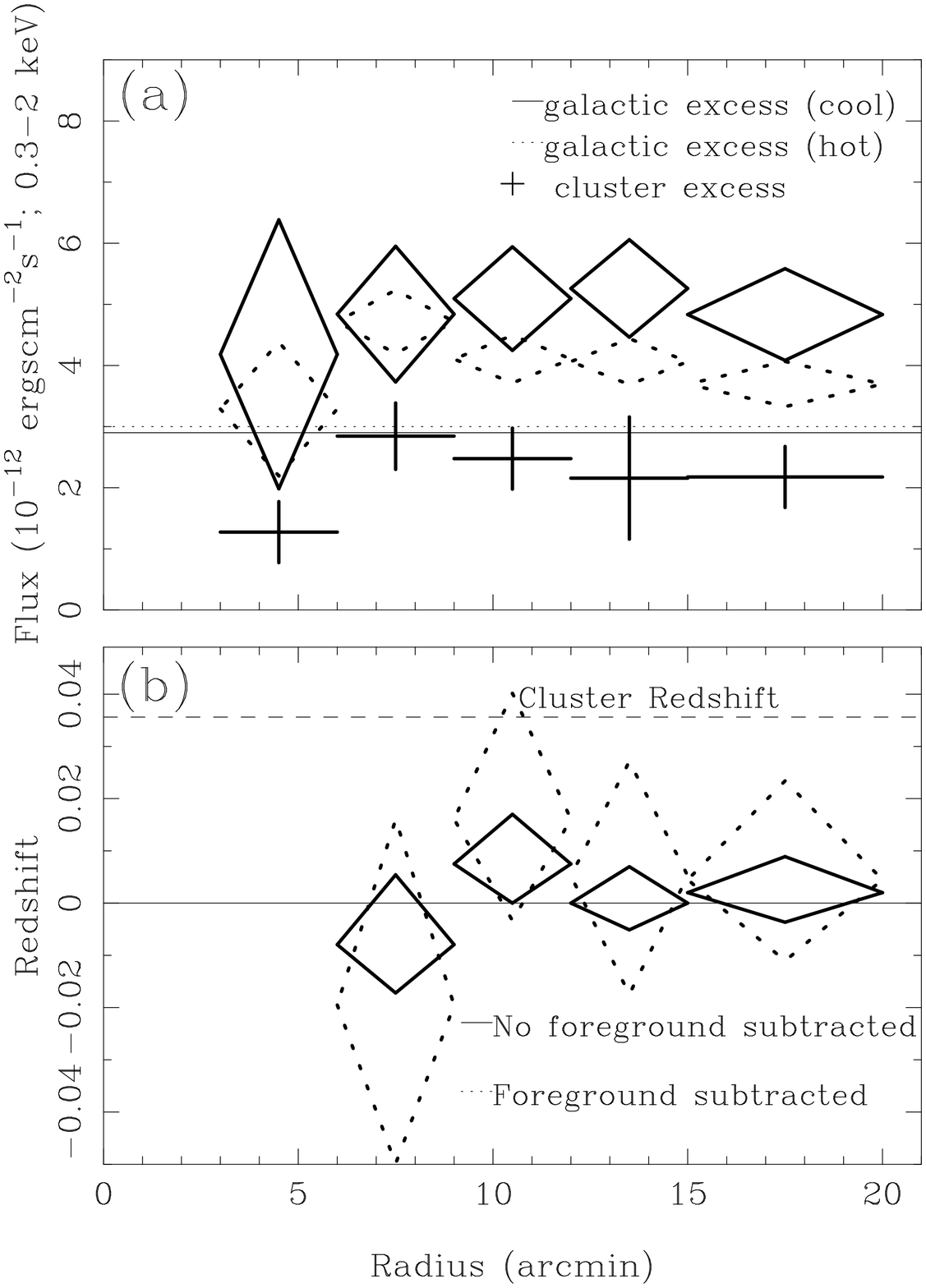,height=0.5\textheight}
\end{center}
  \caption{(a) Radial distributions of the soft component X-ray fluxes in the 0.3--2.0~keV band.
From the galactic excess model, those of the two CIE components, 0.094~keV (black) and 0.34~keV (red), 
are shown in diamonds.
Those from the offset-pointing are shown in two horizontal lines.
Those from the cluster excess model are indicated by cross data points (blue).
Fluxes are scaled to a sky area of 1257 arcmin$^2$ ($20'$ radius circle).

(b) The radial profile of the position of the O\emissiontype{VII} line.
The black- and red-lines indicate redshifts derived from spectra without and with subtracting foreground emission model, 
respectively.

}
\label{fig:flux}
\end{figure}

Next, to describe these residuals, 
we attempt to use the following two models separately.
In the first model,
we assume that the surface brightnesses of the two Galactic foreground components
vary between the offset and cluster regions ('Galactic excess model').
The two temperatures of the Galactic components are fixed to those derived in the offset pointing (0.094~keV and 0.34~keV). 
Therefore additional parameters are two normalizations of Galactic components.
The fitting results are shown in Table~\ref{tbl:fit} and figures~\ref{fig:spec1}-\ref{fig:spec2}.
Except for the inner most region, the fit improved significantly compared with the no excess model
and the model gave a reasonably acceptable description.
As shown in Figure~\ref{fig:flux}(a),
the brightness of the 0.094~keV (0.34~keV) component over the $6'<r<20'$ region 
is about 1.7 (1.4) times larger than that of the offset pointing. 
At $r>3'$, the brightness of each component is radially constant within errors.

In the second model, 
we fixed the Galactic emission to be the same as the offset pointing
and added a thermal component originated from a warm plasma at the cluster redshift
 ($z_{\rm cluster}=0.0356$; 'cluster excess model').
The warm plasma is modeled by 
a single temperature CIE with a fixed metal abundance, 0.1~solar.
Additional parameters are the temperature and normalization of the warm plasma.
Similarly to the Galactic excess model above, 
this model gave improved fits in all radial regions except for the inner most bin (Table~\ref{tbl:fit} and figures~\ref{fig:spec1}-\ref{fig:spec2}).
The temperatures of the excess component are 0.15--0.25~keV.
As shown in Fig.~\ref{fig:flux}(a), 
the X-ray brightness of  the excess component are almost radially constant over the cluster region ($r>6'$).

\begin{table*}[p]
  \caption{Spectral fitting results for the cluster regions.}
\label{tbl:fit}
  \begin{center}
    \begin{tabular}{llcccccc}
\hline \hline
\multicolumn{2}{l}{Projected Radius} & $3'-6'$ & $6'-9'$ & $9'-12$ & $12'-15'$ & $15'-20'$ & sum($6'-20'$)\footnotemark[$*$]  \\
\hline
\multicolumn{8}{c} {No excess model\footnotemark[$\dagger$] (ICM+OFFSET\footnotemark[$\ddagger$]+CXB)} \\
ICM & kT (keV)            & 3.0  & 2.8 & 1.9 & 1.3 & 0.78 & --\\
    & Metal (solar)       & 0.42 & 0.24 & 0.05 & 0.0 & 0.0 & --\\
    & $\chi^2$            & 243  & 263 & 278 & 264 & 225 & 1030 \\
    & d.o.f.              & 165 & 165 & 145 & 129 & 129 & 568 \\
\hline
\multicolumn{8}{c} {Galactic excess model\footnotemark[$\S$] (ICM+Excess1+Excess2+CXB)} \\ 
ICM & kT (keV)            & 3.1$\pm0.05$ & 3.0$\pm0.1$  & 2.3$\pm0.08$  & 2.0$\pm0.2$  & 1.4$\pm0.2$  & --\\
    & Metal (solar)       & 0.43$\pm0.03$  & 0.32$\pm0.04$  & 0.15$\pm0.05$  & 0.13$\pm0.05$  & 0.09$\pm0.06$  & --\\
 & $\chi^2$            & 242 & 203 & 201 & 184 & 197 & 785 \\
 & d.o.f.              & 163 & 163 & 143 & 127 & 127 & 560 \\
\hline
\multicolumn{8}{c} {Cluster excess model (ICM+Excess1+OFFSET\footnotemark[$\ddagger$]+CXB)} \\
ICM & kT (keV)            & 3.1$\pm0.03$ & 3.0$\pm0.1$ & 2.1$\pm0.1$ & 1.7$\pm0.1$ & 1.2$\pm0.1$ & -- \\
    & Metal (solar)       & 0.43$\pm 0.04$ & 0.29$\pm 0.04$ & 0.11$\pm0.03$  & 0.06$\pm0.03$  & 0.07$\pm0.03$ & -- \\
Excess1 &  kT (keV)            & 0.15$\pm 0.08$ & 0.25$\pm0.03$ & 0.19$\pm0.02$  & 0.19$\pm0.02$  & 0.18$\pm0.02$ & -- \\
& $\chi^2$            & 244 & 203 & 194 & 202 & 167 & 766  \\
& d.o.f.              & 163 & 163 & 143 & 127 & 127 & 560 \\
\hline
\multicolumn{8}{c} {Galactic excess model (ICM+Excess1+Excess2+CXB)} \\ %
M1  & $\chi^2$/d.o.f.            & 239/161 & 198/161 & 191/141 & 178/125 & 166/125 & 733/552 \\ 
M2a (CXB x 1.2) & $\chi^2$/d.o.f.            & 242/163 & 201/163 & 196/143 & 183/127 & 210/127 & 790/560 \\ 
M2b (CXB x 0.8) & $\chi^2$/d.o.f.            & 243/163 & 205/163 & 205/143 & 186/127 & 197/127 & 793/560 \\
M3 (E$>0.7$~keV) & $\chi^2$/d.o.f.    & 157/138 & 155/138 & 147/118 & 130/102 & 98/102 & 530/460 \\ 
\hline
\multicolumn{8}{l}{ \parbox{170mm}{\footnotesize
\footnotemark[$*$] Total $\chi^2$ and d.o.f values over $6'-20'$ regions.\\
\footnotemark[$\dagger$] No fitting error is given, because fits are significantly poor.\\
\footnotemark[$\ddagger$] The Galactic foreground emission model derived in \S~\ref{sect:offset} (2CIE model).\\

}}
    \end{tabular}
  \end{center}
\end{table*}

\subsection{The O\emissiontype{VII} Line Position of the Excess Emission}
We examine the excess spectra in the outer part of the cluster ($r>6'$) for the position of the O\emissiontype{VII} line.
Because of a larger effective area, the oxygen line can be seen more clearly in the BI spectra than in the FI one.
Therefore, we use only the BI spectrum in the limited energy range around the line (0.50--0.65~keV).
Note that the reported calibration uncertainty of the line energy at energies below $1$~keV is about 5--10~eV\footnote{http://www.astro.isas.jaxa.jp/suzaku/process/caveats/}.
This corresponds to redshift of 0.01--0.02 at the oxygen line energy.

Firstly, we examine the spectrum without subtracting any foreground emission.
In this case, the O\emissiontype{VII} line is modeled by a 0.1~keV CIE component as used in the previous subsections (e.g. No excess model ).
In Fig.~\ref{fig:flux}(b) we show a radial profile of the best-fit redshift of the CIE component.
The redshifts are consistent with being radially constant at zero.
The redshift of the spectrum integrated over $9'-20'$ region is $0.004\pm0.004$~(statistical error). 
Even if combining this error with the systematic one, 
we can reject that all the O\emissiontype{VII} line emission originates from the cluster redshift at $0.0356$.

Secondly, we subtract the foreground emission model derived from the offset pointing (\S~\ref{sect:offset}) and examine the residual O\emissiontype{VII} line emission.
The residual emission is modeled by an additional Gaussian line with a zero intrinsic width.
The derived line energies are converted to redshift assuming intrinsic line energy of 569~eV.
The positions of the residual lines [Fig.~\ref{fig:flux}(b)] are closer to zero than to the cluster redshift.
The redshift of the residual spectrum 
integrated over $9'-20'$ region is $0.009\pm 0.012$. 
The average O\emissiontype{VII} line flux is 4.1~LU.
The systematic uncertainty of the redshift in this method 
is approximately similar to that of the absolute energy scale (i.e. 0.01--0.02 in redshift).
Therefore, 
combining these statistical and systematic errors, 
we conclude that the data are consistent with either cluster ($z=0.0356$) or Galactic ($z=0$) origins of the residual emission.

\subsection{The ICM Radial Properties} \label{sect:icm-radial}
\begin{figure*}[htbp]
\begin{center}
\leavevmode\psfig{figure=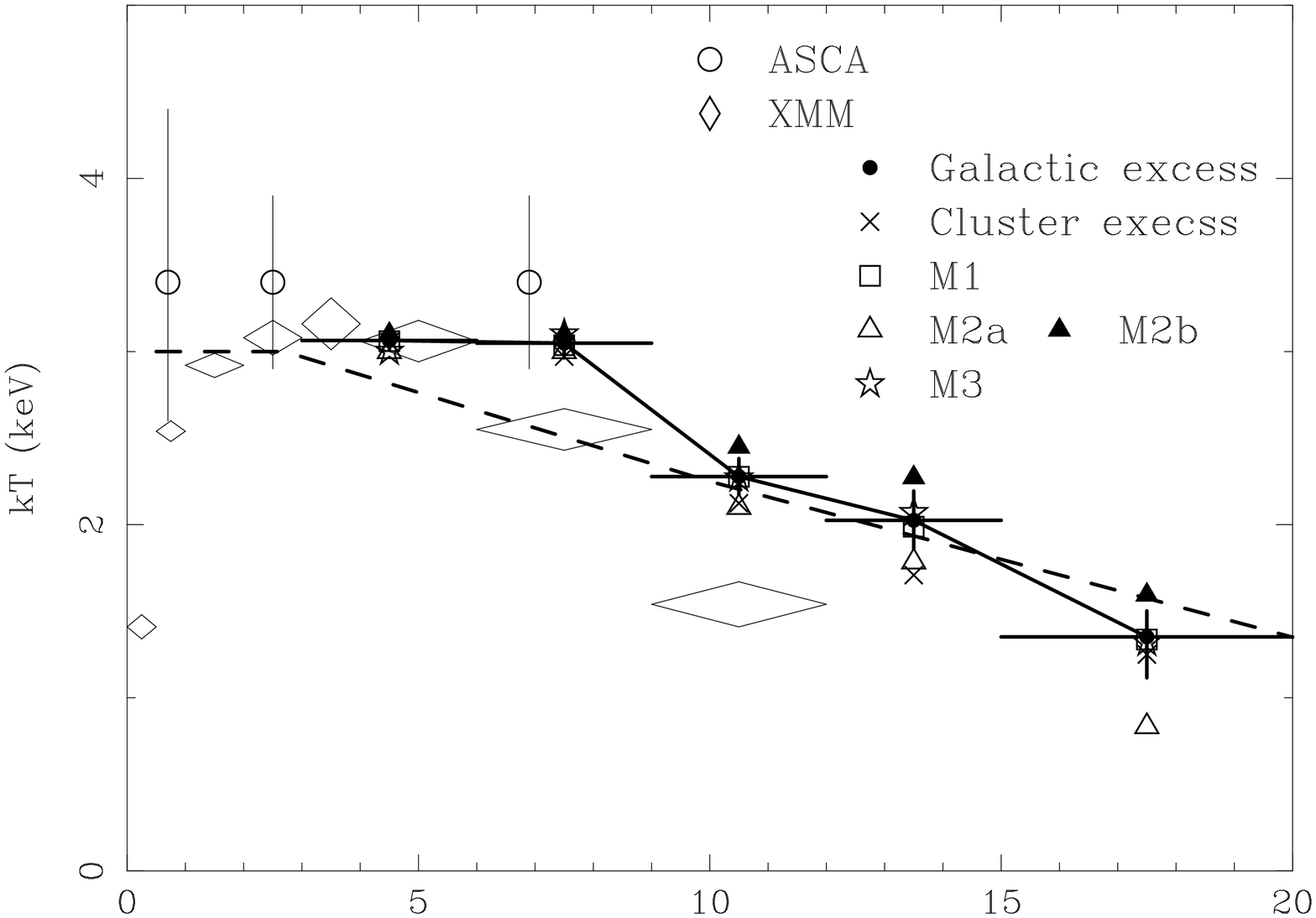,angle=-90,width=0.688\textwidth}
\leavevmode\psfig{figure=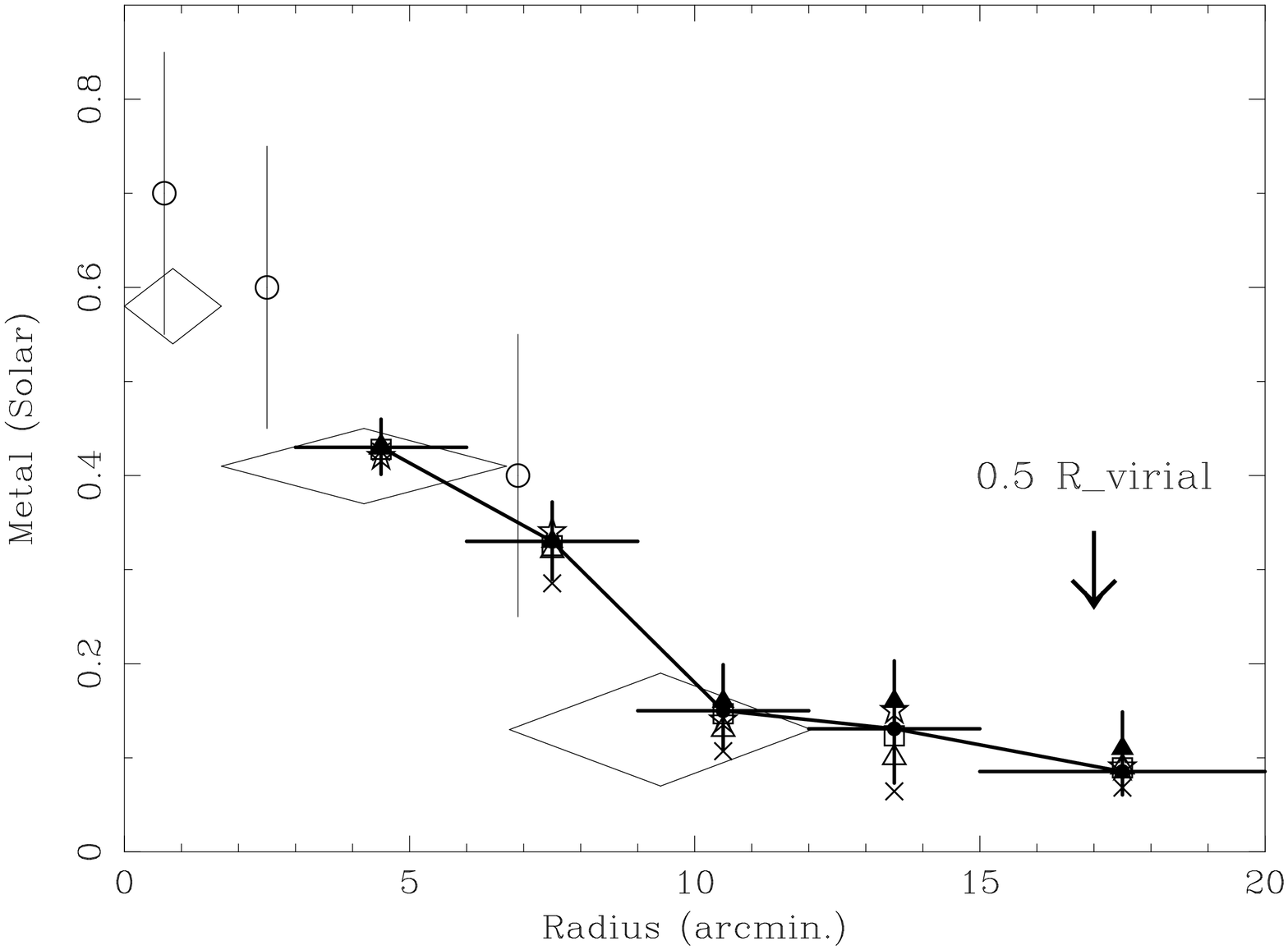,angle=-90,width=0.7\textwidth}
\end{center}
  \caption{Radial profiles of the ICM temperature (upper panel) and metal abundance (lower panel).
Results from different modeling are shown by different marks.
Errors are given only from the Galactic excess model. 
Errors from other models are approximately similar to those from the Galactic excess model. 
Previous results from {\it ASCA} (Finoguenov et al. 2001) and {\it XMM-Newton} (Kaastra et al. 2004; Tamura et al. 2004) 
are shown by open-circles and diamonds, respectively.
In the upper panel, the universal temperature profile derived from a {\it Chandra} study (Vikhlinin et al. 2005) by scaling for A~2052 are shown by dashed-line.
In the lower panel, the position of half of the virial radius ($17'$) is indicated.
}
\label{fig:r-kt}
\end{figure*}

In \S~\ref{sect:cluster-spectra} above, 
we obtained acceptable descriptions of the observed spectra 
by models consisting of the ICM and Galactic or cluster excess emission along with other backgrounds.
Using these models,
we constrain the radial ICM temperature and metal abundance as shown in Figure~\ref{fig:r-kt}.
The two models give results almost consistent with each other.

To further investigate possible uncertainties of derived parameters,
we attempt to use different fitting and modeling methods as follows.
In all cases, we started from the Galactic excess model.
The best-fit parameters and $\chi^2$ values from these methods are presented in Fig.~\ref{fig:r-kt}
and Table~\ref{tbl:fit}, respectively.
In the first case (M1), 
we allow the two temperatures of the Galactic components to vary within the uncertainty in the offset fitting,
$0.084 < kT_{\rm 1}$ (keV) $ < 0.104$ and $0.32 < kT_{\rm 2}$ (keV) $ < 0.36$.
This model gives better fit to the data compared to the original Galactic excess model.
In the second case (M2a, M2b), 
to understand effects of CXB fluctuations,
we change the CXB normalization by $\pm$20\% from the standard value which is used in other models.
This variation is approximated from the observed fluctuation by Kushino et al. (2002), 
who reported a standard deviation of 6.5\% for a 0.5 deg$^2$ effective area.
In the third case (M3), 
to avoid effects of the soft excess emission as much as possible,
we ignore the energy range below 0.7~keV. 

As seen in Fig.~\ref{fig:r-kt}, all methods give results consistent within statistical errors 
with those obtained in \S~\ref{sect:cluster-spectra}, 
except for the outermost bin.
At the $15'-20'$ region, 
temperature depends on the assumed CXB normalization largely than the statistical uncertainty.
On the other hand, modeling of the soft excess components
gives little uncertainty.
Regardless of methods, both ICM temperature and metallicity show a radial decline.


\section{Discussion}
\subsection{Origins of the Soft X-ray Excess}
\subsubsection{Results}
Focusing on soft X-ray spectral properties, 
we analyzed {\it Suzaku} data of the A~2052 cluster and offset (\timeform{3D.74}) regions.
We summarize results as follows.
(1) The offset spectrum already shows excess emission at energies below and above the O\emissiontype{VII} line position, as compared with another blank-sky field (NEP).
(2) Assuming this offset spectrum as the Galactic foreground emission and removing other background components and the ICM emission, we confirmed significant soft X-ray excess emission in the direction to the A~2052.
(3) This excess can be described either by
(a) a combination of increases of the two Galactic thermal components ($\sim$0.1~keV and $\sim$0.3~keV) or
(b) an additional thermal component with a temperature of about 0.2~keV.
(4) Regardless of the modeling uncertainty, 
spectral properties and flux of the excess emission are radially uniform 
at least at the outer part of the cluster ($6'<r<20'$).
(5) The data suggest that the O\emissiontype{VII} line emission after subtracting the Galactic foreground model 
comes from zero redshift rather than from the cluster one,
although the cluster origin can not reject conservatively.
We discuss possible origins of the soft excess in the offset and cluster regions below.

\subsubsection{Galactic Origins}
As stated previously, the A~2052 and offset pointing positions are at a tips of the NPS.
Recent {\it XMM-Newton} and {\it Suzaku} observations revealed that the NPS emission can be described by a thermal spectrum with a temperature of $0.25-0.3$~keV (Willingale et al. 2003; Miller et al. 2008).
This spectral feature is similar to what we found from the offset pointing in comparison with another blank sky field.
Therefore, the excess in the offset region is most likely associated with the NPS.
We presume further that 
variation of the NPS emission 
causes at least the enhancement in the 0.3~keV component in the cluster direction.
The emission measure of the 0.3~keV component from the offset and cluster positions 
are about 25\% and 35\% of that from the brightest position in the NPS derived by Miller et al. (2008).

The O\emissiontype{VII} line flux in the offset pointing shows an insignificant excess compared with other blank fields.
On the other hand,
the 0.1~keV component which is dominated by the oxygen line
in the cluster direction shows a clear enhancement above that of the offset pointing data.
This kind of spectral component is usually attributed to 
the local bubble near the sun or more distant Galactic halo.
The flux in this component is 70\% higher in the cluster region than in the offset region.
Willingale et al. (2003) found a similar level of variation in the 0.1~keV component within the NPS region ($l=$\timeform{20D-25D}, $b=$\timeform{20D-40D}).
In addition,
from several {\it ROSAT} pointing observations in the surrounding area of A~2052,  
Bonamente et al. (2005) reported a standard deviation of 81\% in the diffuse X-ray flux of the R4 band ($\sim0.4-1.0$~keV) including the oxygen line.
Note that in their calculation, data from cluster regions are excluded.
Therefore the 70\% increase of the 0.1~keV component observed here
is possibly caused by variation of the Galactic foreground components.

\subsubsection{Cluster Origins}
Using the cluster excess model in \S~\ref{sect:cluster-spectra}, 
we found that the data are consistent with that the excess originating from a warm plasma at the cluster distance.
Using a similar spectral modeling, Kaastra et al. (2003a; 2003b) analyzed the {\it XMM-Newton} data.
Here we compare the two results.
Note that while Kaastra et al. observed the cluster within $12'$ in radius,
we have extended the region up to almost $20'$.
Assumed and derived temperatures and metal abundances of the two studies  
are the same ($\sim 0.2$~keV and 0.1~solar).
Both results consistently show a radially uniform brightness distribution of the warm plasma emission.
However, we found that the brightness reported in Kaastra et al. (2003b) in terms of the emission measure per solid-angle [$(7.3\pm 1.5)\times 10^{69}$~m$^{-3}$arcmin$^{-2}$] is 
higher by a factor of two than our result.
In our case, spectral normalizations from the cluster excess model in \S~\ref{sect:cluster-spectra} give an emission measure averaged over $6'<r<20'$ region of $3.7\times 10^{69}$~m$^{-3}$arcmin$^{-2}$.
This discrepancy could be due to largely a difference in the foreground emission estimation.
Kaastra et al. used a sky-average foreground.
Based on the {\it Suzaku} offset data, 
we found that the foreground in this area is enhanced at least in the energy range of 0.6--0.9~keV,
with respect to the NEP region and hence probably to the sky average.
Note that a 0.2~keV thermal emission peaks around this energy range and hence its brightness is determined mostly in this range.
Therefore, we presume that Kaastra et al.(2003a; 2003b) underestimated the foreground emission.

The possibility of the warm plasma associated with the cluster and/or larger scale structure
is already discussed by Kaastra et al. (2003a) and Bonamente et al. (2005) in this cluster.
Consistently with Kaastra et al. (2003a) we found lack of radial variations 
in the spectral feature and brightness of the excess emission within the observed region.
This is a contrast to the galaxy or ICM density distributions, 
each of which shows a clear central increase.
In other words, 
there is no correlation between the soft excess emission and the galaxy or ICM distribution.
Furthermore, the spectral shape of the excess emission is similar to that of the Galactic foreground emission.
Therefore it is difficult to associate this emission with the cluster.
If we presume that observed excess originates from a uniform density warm plasma around the cluster
and electron number density to be 1.2 times the hydrogen density ($n_{\rm H}$), 
the observed emission measure given above
can be translated to $n_{\rm H}$ as
\begin{equation}
 n_\mathrm{H} = 1.7\times10^{-4}~\mathrm{cm^{-3}}~
  \left(\frac{Z}{0.1~Z_\mathrm{\odot}}\right)^{-1/2}~
  \left(\frac{L}{2~\mathrm{Mpc}}\right)^{-1/2},
\end{equation}
where $Z$ is the oxygen abundance and $L$ is the path length.
Note that we assume also that the derived emission measure is proportional to $Z^{-1}$ since the emission is dominated by the the oxygen line.
Note also that $L=2$~Mpc corresponds to $48'$, approximately the diameter of the observed region.
This density is a few times higher than 
upper limits reported from {\it Suzaku} observations of other clusters (
$0.78\times10^{-4}~\mathrm{cm^{-3}}$; Takei et al. 2007, 
$0.41\times10^{-4}~\mathrm{cm^{-3}}$; Fujita et al. 2007).
Therefore it is unlikely that all the excess emission originates from the cluster region.

\subsection{The ICM Radial Properties}
Focusing on the ICM properties, we discuss systematic uncertainties of the measurement
and compare our results with previous ones.
Then we point out importance of the measurement.

The spatial resolution of {\it Suzaku} is lower than those of {\it XMM-Newton} or {\it Chandra}.
This may cause systematic errors on the ICM radial properties.
The {\it Suzaku} point spread function has a half power diameter of about $2'$, 
which is smaller than the radial bin size used here.
In addition, the point spread function has negligible dependence on the photon energy.
Sato et al. (2007) made a ray-tracing telescope simulation for a cluster (A~1060), 
which has a radial brightness profile similar with A~2052 in the outer region.
Based on their result, 
we estimate that more than 65\% of the photons originates from the corresponding sky annulus at least in the outer region ($r>6'$).
Therefore we approximate that the error caused by this spectral mixing
is small compared with the statistical one.

We examined uncertainties of the ICM properties caused by the CXB fluctuation in \S~\ref{sect:icm-radial}. 
The instrumental background count is a few times lower
and the level of its uncertainty is not larger than those of the CXB
over the most part of the detector.
Therefore the error caused by the instrumental background uncertainty
should be smaller than that by the CXB variation.

Here we compare derived properties with previous measurements (Fig.~\ref{fig:r-kt}).
Finoguenov et al. (2001) reported temperature and metallicity profiles of the cluster at $r<8'$ 
using {\it ASCA}.
Their temperature of $3.5\pm 0.5$~keV is slightly higher than our result.
Using {\it XMM-Newton}, Kaastra et al. (2004) reported the temperature profile within $12'$
focusing on the cluster core.
Although the temperature at the inner region ($1'<r<6'$) is consistent with our result,
their result shows a steeper decline beyond that radius than our result.
This difference is most likely caused by that 
Kaastra et al. (2004) ignored the modeling of the soft excess emission in this cluster.
The metallicity within $12'$ derived from {\it XMM-Newton} by Tamura et al. (2004) is consistent with ours.
We could measure the ICM properties beyond regions by previous observations in this cluster for the first time.
The obtained temperature profile is consistent with the universal temperature profile derived from a sample of cluster by {\it Chandra} (Vikhlinin et al. 2005; Fig.~\ref{fig:r-kt}).
Here we scaled their profile to A~2052 with a peak temperature of 3.0~keV and a virial radius of $34'$ (1.4~Mpc).

A temperature profile is crucial to determine the total mass of a cluster based on the hydrostatic equilibrium assumption.
For example,
our derived temperature profile gives the total mass of $10^{14}$~\MO within $15'.5$.
Here we assume the gas density profile at that radius to be proportional to $R^{-3\beta}$ with $\beta=0.7$ (Mohr et al. 1999), where $R$ is the radius.
On the other hand, a 3~keV isothermal profile gives 50\% higher mass within the same volume.

Compared with the ICM temperature,
the metal abundance at cluster outer regions have been much more difficult to constrain.
Except for a few cases, 
previous observations are limited to the cluster inner region (smaller than 0.25--0.4 of the virial radius).
Our result gives a gas-mass-weighted metal abundance averaged over the observed region ($r<20'$) 
to be $0.21\pm 0.05$ times solar. 
We found a large scale abundance gradient, 
from 0.4~solar at the inner region to 0.1~solar beyond $r = 15'$ (about half the virial radius).
The latter value is significantly smaller than
previously measured 'cluster  values' of 0.3--0.4 (Fukazawa et al. 1998, De Grandi et al. 2004, Tamura et al. 2004).
A similar kind of the abundance profile has been found in AMW~7 (Ezawa et al. 1997) and the Perseus cluster (Ezawa et al. 2001).
As argued by Metzler \& Evrard (1994),
these gradients could be caused by steeper distribution of galaxies (origin of metals) than the ICM.
These also suggest that no strong mixing in the ICM after the metal injection.
In addition,
a radial change of efficiencies of different enrichment processes could affect the abundance profile.
In the inner region where the ICM density is high ram-pressure stripping should be effective, 
while in the outer region galactic wind could dominate the metal pollution.
Kapferer et al. (2007) confirmed this trend quantitatively by simulation with a semi-numerical galaxy formation model.

On the contrary to these gradients above, 
Fujita et al. (2008) found an uniform abundance ($\sim 0.2$~solar) upto the virial radius in the link region between two clusters, A~399 and A~401.
Based on the uniformity, they argued a possibility that the proto-cluster region was already metal enriched by past galactic superwinds.
Note that we and Ezawa et al. (1997; 2001) measured azimuthal averaged abundances, 
while Fujita et al. (2008) measured a local region, a possible cosmic filament.
These measurements of the large scale abundance profile and
precise metal mass in the ICM along with the abundance ratios
are of importance to constrain origins, enrichments, and transport of the metals
and in turn past dynamical history of clusters.

\bigskip
We thank anonymous referee for useful comments.
JPH is supported by NASA Grant NNG06GC04G.
We thank all the {\it Suzaku} team member for their supports.



\begin{thebibliography}{31}
\bibitem[{Anders \& Grevesse(1989)}]{anders89:_abund}
Anders, E., \& Grevesse, N. 1989, Geochimica et Cosmochimica Acta, 53, 197 
\bibitem[{{Bonamente} {et~al.}(2005){Bonamente}, {Lieu}, \&
  {Kaastra}}]{2005A&A...443...29B} 
{Bonamente}, M., {Lieu}, R., \& {Kaastra}, J. 2005, \aap, 443, 29 
\bibitem[]{}
Bregman 2007, \araa, 45, 221, 
\bibitem[]{}
Cen, R., \& Ostriker 1999, \apj, 514, 1 
\bibitem[]{}
De Grandi, S. Ettori, S., Longhetti, M. \& Molendi, S. 2004, \aap, 419, 7 
\bibitem[{Dickey \& Lockman(1990)}]{dickey90}
Dickey, J.~M., \& Lockman, F.~J. 1990, ARAA, 28, 215
\bibitem[]{}
Ezawa, H., Fukazawa, Y., Makishima, K., Ohashi, T., Takahara, F., Xu, H., \& Yamasaki, N.Y. 1997, \apj, 490, L33 
\bibitem[]{}
Ezawa, H., et al. 2001, \pasj, 53, 595 
\bibitem[]{}
Egger, R.J. \& Aschenbach, B. 1995, \aap, 294, L25 
\bibitem[]{}
Fukazawa et al. 1998, \pasj, 50, 187 
\bibitem[]{}
Finoguenov, A., Arnaud, M., \& Dadid, L.P. 2001, \apj, 555, 191 
\bibitem[]{}
Fujimoto, R., 2007,  \pasj, 59, S133 
\bibitem[]{}
Fujita, Y., 2008,  \pasj, 60, S343 
\bibitem[]{}
Ishisaki, Y., et al. 2007,  \pasj, 59, S113 
\bibitem[{{Kaastra} {et~al.}(2003a){Kaastra}, {Lieu}, {Tamura}, {Paerels}, \&
{den Herder}}]{2003A&A...397..445K} 
{Kaastra}, J.~S., {Lieu}, R., {Tamura}, T., {Paerels}, F.~B.~S., \& {den  Herder}, J.~W. 2003, \aap, 397, 445 
\bibitem[]{}
Kaastra, J.~S., Lieu, R., Tamura, T., Paerels, F.B.S., \& den Herder, J.W., 2003b,  
in conference ``Soft X-ray emission from clusters of galaxies and related phenomena'' (astro-ph/0305424) 
\bibitem[]{}
{Kaastra}, J.~S., et al. 2004, \aap, 413, 415 
\bibitem[]{}
Kapferer, W., et al. 2007, \aap, 466, 813 
\bibitem[Koyama et~al. (2007)]{Koyama2007}
Koyama, K., et al. 2007,  \pasj, 59, S23 
\bibitem[{{Kushino} {et~al.}(2002){Kushino}, {Ishisaki}, {Morita}, {Yamasaki},
  {Ishida}, {Ohashi}, \& {Ueda}}]{2002PASJ...54..327K}
{Kushino}, A., {Ishisaki}, Y., {Morita}, U., {Yamasaki}, N.~Y., {Ishida}, M.,
  {Ohashi}, T., \& {Ueda}, Y. 2002, \pasj, 54, 327
\bibitem[]{}
Metzler, C.A., \& Evrard, A.E., 1994, \apj, 437, 564
\bibitem[]{}
Miller, E., et al. 2008,  \pasj, 60, S95
\bibitem[Mitsuda et~al. (2007)]{Mitsuda2007}
Mitsuda, K., et al. 2007,  \pasj, 59, S1 
\bibitem[]{}
Mohr, J.J., Mathiesen, B., \& Evrard, A.E. 1999, \apj, 517, 627 
\bibitem[]{}
Sato, K. et al. 2007, \pasj, 59, 299 
\bibitem[Serlemitsos et al. 2007]{}
Serlemitsos, P. et al., et al. 2007,  \pasj, 59, S9 
\bibitem[]{}
Snoden, S.L.et al. 1995, \apj, 454, 643 
\bibitem[]{}
Smith, R.K \& Brickhouse, N.S. 2001, \apj, 556, L91 
\bibitem[]{}
Smith, R.K et al. 2007, \pasj, 59, S141 
\bibitem[]{}
Takei et al. 2007, \pasj, 59, S339 
\bibitem[{{Tamura} {et~al.}(2004){Tamura}, {Kaastra}, {den Herder}, {Bleeker},
  \& {Peterson}}]{2004A&A...420..135T}
{Tamura}, T., {Kaastra}, J.~S., {den Herder}, J.~W.~A., {Bleeker}, J.~A.~M., \&
  {Peterson}, J.~R. 2004, \aap, 420, 135 
\bibitem[]{}
Tawa, N., 2008,  \pasj, 60, S11
\bibitem[]{}
Vikhlinin, A., Markevitch, M., Murray, S.S., Jones, C., Forman, W., \& Van Speybroeck, L.
2005, \apj, 628, 655 
\bibitem[]{}
Willingale, R., Hands, A.D.P., Warwick, R.S., Snowden, S.L., \& Burrows, D.N. 2003, \aap, 343, 995 

\end{thebibliography}
\end{document}